\documentclass[book,oneside,11pt]{article}

\newtheorem{dfn}{Definition}[section]
\newtheorem{tw}[dfn]{Theorem}
\newtheorem{prop}[dfn]{Proposition}

\newtheorem{lem}[dfn]{Lemma}

\usepackage{amssymb} 
\usepackage{amsmath}
\numberwithin{equation}{section}

\usepackage{anysize,color}
\usepackage{array}
\usepackage{enumerate}

\let\g\gamma

\author{Micha\l \ Barski \\ \small  Faculty of Mathematics, Cardinal Stefan Wyszy\'nski University in Warsaw, Poland\\
\small Faculty of Mathematics and Computer Science, University of Leipzig, Germany\\ \small{\it Michal.Barski@math.uni-leipzig.de} \bigskip \\
\\
Jerzy Zabczyk
\\ \small Institute of Mathematics, Polish Academy of
Sciences,
     Warsaw,  Poland\\ \small{\it zabczyk@impan.pl}}

\title{\bf Heath-Jarrow-Morton-Musiela  equation   with L\'evy perturbation\footnote{Supported by The Polish MNiSW grant
NN201419039.}}

\begin{document}
\baselineskip=1.2\baselineskip \maketitle
\date
\begin{abstract}
The paper studies  the Heath-Jarrow-Morton-Musiela equation of the bond market. The equation is analyzed  in weighted spaces of functions defined on
$[0,+\infty)$. Sufficient conditions for local and global existence are obtained . For equation with the linear diffusion term the conditions for global
existence are close to the necessary ones.
\end{abstract}

\tableofcontents

\section{Introduction}
The Heath-Jarrow-Morton-Musiela equation, driven by a real L\'evy process $L$,   is a stochastic partial
differential equation of the form
\begin{align}\label{general equation}\nonumber
dr(t,x)=&\left[\frac{d}{dx}r(t,x)+ F(r(t))(x)\right]dt+G(r(t-))(x)dL(t),\\[2ex]
r(0,x)=&r_{0}(x),\,\,\,\,\,\,x\geq 0,\,\,\,\,t\in (0,T^\ast ],
\end{align}
where the diffusion operator $G$ and the drift $F$ are of the form:
\begin{gather}\label{volatility and drift}
 G(r)(x):=g(x,r(x)); \qquad F(r)(x):=J^{\prime}\left(\int_{0}^{x}g(v,r(v))dv\right)g(x,r(x)).
\end{gather}
The function $J^{\prime}$ admits a representation
\begin{equation}\nonumber
J^{\prime}(z)= -a + qz +\int_{\mathbb{R}}y(\mathbf{1}_{(-1, 1)}(y)-e^{-zy}) \, \nu(dy),\qquad z\in\mathbb{R},
\end{equation}
with $ a\in \mathbb{R}$, $q\geq 0$ and the measure $\nu$ satisfies
the following integrability condition
\begin{equation}\nonumber
\int_{\mathbb{R}}(y^2\wedge \ 1) \ \nu(dy)<\infty.
\end{equation}
The measure $\nu$ is the L\'evy measure of the process
$L$. The function $g$ has a financial meaning and is sometimes called volatility
of the bond market. Solutions to \eqref{general equation} are, the so called, forward curves, see e.g.
\cite{Filipovic}, \cite{Filipovic1}, \cite{PeszatZabczyk} and the quantity
\begin{gather*}
P(t,T)=e^{-\int_{0}^{T-t}r(t,v)dv}, \qquad t\leq T,
\end{gather*}
can be interpreted as the price, at moment $t$, of the bond which matures at moment $T$
(and then pays $1$). {The equation (\ref{general equation}) describes the dynamics of the forward curves in the {\it moving frame} and was introduced by Musiela in \cite{Musiela}. The original version, in the {\it natural frame}, appeared first in the PhD dissertation of Morton \cite{Morton}. For more information about the financial background of the equation}  see Appendix \ref{HJM approach to the bond market}.

If the process  $L$ is not present  in the equation, that is if $L$  is identically zero, then  
$J'=0$, $F=0$ and $G=0$ and the equation has trivial solution $r(t,x)= r_0(t+x)$. So
only the stochastic case is of real interest. The equation was intensively
studied in the case when $L$ is a Wiener process, see e.g.
\cite{Filipovic1}, \cite{PeszatZabczyk} and references therein. Then
the function $J^{\prime}$ is linear: $J^{\prime}(z)= qz,\,\,z\in \mathbb{R},\,\,\,q\geq 0$.
There are also several results for the case of general infinite dimensional L\'evy
process $L$, see e.g. \cite{PeszatZabczyk}, \cite{FilTap},
\cite{Rusinek}, \cite{Rusinek2}, \cite{Rusinek3}, \cite{Marinelli}, \cite{FilipovicTappeTeichmann}, \cite{PeszatZabczyk1}. 
In particular in \cite{Marinelli} local solvability of \eqref{general equation} was studied for L\'evy process $L$ having exponential moments, the assumption which we find very restrictive. In fact a majority of the results  presented in the paper can be extended to infinite dimensional noise. Our intention was to obtain optimal results in the most important case of the one dimensional process $L$,  to see what kind of results could be expected in the general case.

The aim of the present paper is to establish existence and uniqueness of weak solutions to (\ref{general equation}). We restrict our attention to
positive solutions { which are relevant for applications}. The equation is studied either in the Hilbert space $ H=L^{2,\gamma}$, of
square summable functions $h$ on $[0, +\infty)$ with the norm
\begin{equation}\label{L-square}
  \|h\|_{L^{2,\gamma}}:=\left(\int_{0}^{+\infty}\mid h(x)\mid^2e^{\gamma x}dx\right)^{\frac{1}{2}}<+\infty ,
\end{equation}
or in the Hilbert space $H=H^{1,\gamma} $, of absolutely continuous functions $h$ on $[0,+\infty)$ such that
\begin{equation}\label{H square}
\|h\|_{H^{1,\gamma}}:=\left(\int_{0}^{+\infty}\left(\mid
h(x)\mid^2+\mid h^{\prime}(x)\mid^2\right)e^{\gamma x}dx\right)^{\frac{1}{2}}<+\infty,
\end{equation}
with $\gamma >0$. {Similar results can be obtained for spaces with  different weight functions.}\vskip3mm

{The  paper is divided into Part \ref{General case} and
Part \ref{Linear case}}. Part \ref{General case} studies equation \eqref{general equation} with general
volatility $g$ and uses some versions of the contraction mapping
theorem. Part \ref{Linear case}  is devoted to the case when $g$ is a
linear function of the second variable:
\begin{gather*}
g(x,y) = \lambda (x)y, \qquad x,y\geq 0.
\end{gather*}
In the latter case, more special but important, better results can be obtained using some monotonicity properties of the equation.\vskip3mm

Part \ref{General case} starts with formulating local and global
existence results in the sets $L^{2,\gamma}_{+}$ and
$H^{1,\gamma}_{+}$ of positive functions in $L^{2,\gamma}$ and
$H^{1,\gamma}$ respectively, see Theorem \ref{tw o istnieniu
rozwiazan lokalnych w L^2}, Theorem \ref{tw o istnieniu rozwiazan
globalnych w L^2} and Theorem \ref{tw o istnieniu rozwiazan
lokalnych w H^1}, Theorem \ref{tw o istnieniu globalnych rozwiazan w
H^1}. The main tool here is some extension to locally Lipschitz
coefficients of the standard result on existence of positive
solutions to stochastic evolution equations. The proofs start from
establishing first abstract existence results and then proving local
Lipschitz properties and linear growth of the coefficients as well
as checking conditions for positivity. It turns out that only for
restrictive class of functions $J^{\prime}$ the diffusion $G$ and
the drift $F$ can be locally Lipschitz or of linear growth.

Part \ref{Linear case} is devoted to the equation
\begin{align}\label{basic equation}\nonumber
dr(t,x)=&\left[\frac{d}{dx}r(t,x)+ J^{\prime}\left(\int_{0}^{x} \lambda (v)r(t,v)dv\right)\lambda(x)r(t,x)\right]dt +\lambda(x) r(t-,x)dL(t),\\[2ex]
r(0,x)=&r_{0}(x),\,\,\,\,\,\,x\geq 0,\,\,\,\,t\in (0,T^\ast ].
\end{align}
where $ \lambda(\cdot)$ is a {continuous, positive and bounded
function.}   From results of Part \ref{General case} one deduces
easily sufficient conditions for existence of local, positive
solutions to \eqref{basic equation}. They are formulated as Theorem
\ref{local1} and Theorem \ref{local2}. Main results on existence of
global solutions are presented as Theorem \ref{tw o eksplozjach} and
Theorem \ref{tw o istnieniu}. Uniqueness is proved in Theorem
\ref{tw o jedynosci}. Moreover, Theorem \ref{tw o silnym
rozwiazaniu} gives conditions under which global solutions are
strong. The proofs of those results exploit the fact that the weak
form of the equation \eqref{basic equation} is equivalent to the
equation
\begin{gather}\label{rownanie na r}
r(t,x)=a(t,x)e^{\int_{0}^{t}J^{\prime}(\int_{0}^{t-s+x}\lambda(v)r(s,v)dv)\lambda(t-s+x)ds},
\quad x\geq 0,\,\,\,\,t\in (0,T^\ast ],
\end{gather}
where
\begin{align*}
 a(t,x):=&r_0(t+x)e^{\int_{0}^{t}\lambda(t-s+x)dL(s)-\frac{q^2}{2}\int_{0}^{t}\lambda^2(t-s+x)ds}\\[2ex]
&\cdot\prod_{0\leq s\leq t}\left(1+\lambda(t-s+x)\triangle
L(s)\right)e^{-\lambda(t-s+x)\triangle L(s)}.
\end{align*}
The equivalence of \eqref{rownanie na r} and the weak form of \eqref{basic equation} is established in Section  \ref{Formulation of the main results}, in Theorem \ref{tw semigroup form implies operator form} and Theorem \ref{prop o rownowaznoesci rozwiazan},  preceding the proofs of the main
results. {The proofs are rather involved and require some new results on regularity of L\'evy fields of independent interest, see Proposition \ref{prop o
calkowaniu przez czesci}  Proposition \ref{prop o I_2} .}  Standard methods exploiting the Lipschitzianity of the coefficients, applied for instance in
\cite{FilipovicTappeTeichmann}, \cite{Marinelli}, require more restrictive conditions.\vskip2mm

As we have already said, the study of the linear HJMM equation was initiated by Morton, in
his PhD dissertation \cite{Morton}. He showed that {the equation \eqref{basic equation} in the {\it natural frame}}, with $L$ being a Wiener process, does not have a solution in the class of
bounded functions of two arguments on a finite domain. The situation
changes substantially when $L$ is a general L\'evy
process and results on  existence and explosions  for {the equation \eqref{basic equation} but in the {\it natural frame}} were obtained \cite{BarZab}.\vskip2mm

{The present paper is a much elaborated  version of the note \cite{BarZab2} presented in arxiv.}

\vskip2mm
{{\bf Acknowledgements.} The authors would like to thank S. Peszat and A. Rusinek for inspiring discussions on the subject of the paper.}

\section{Preliminaries}

 We gather first results on properties of the {\it Laplace exponent} $J$ of the process $L$ and its derivatives,  which will be frequently used in
 the following sections of the paper. The first derivative $J^{\prime}$, appears explicitly in the basic equation
 (\ref{general equation})-(\ref{volatility and drift}). As our prime issue will be the solvability of (\ref{general equation})-(\ref{volatility and drift})
 in the set of non-negative functions we concentrate on the properties of $J$  and its derivatives for non-negative arguments.\vskip2mm

\noindent  The function $J$ is defined by
\begin{equation}\label{J}
\mathbf{E}(e^{-zL(t)})=e^{tJ(z)},\qquad t\in[0,T^\ast], \ z\in \mathbb{R},
\end{equation}
and admits explicit representation
\begin{equation}\label{Laplace transform}
J(z)=-az+\frac{1}{2}qz^2+\int_{\mathbb{R}}(e^{-zy}-1+zy\mathbf{1}_{(-1, 1)}(y)) \ \nu(dy),\qquad z\in\mathbb{R},
\end{equation}
with $ a\in \mathbb{R}$, $q\geq 0$ and the measure $\nu$ satisfies the following integrability condition
\begin{equation}\label{K1}
\int_{\mathbb{R}}(y^2\wedge \ 1) \ \nu(dy)<\infty.
\end{equation}
It is easy to see that $J$ is well defined for all positive numbers $z$ if and only if
$$
\int_{- \infty}^{-1}e^{z|y|} \nu(dy)<+\infty, \qquad z\geq 0.
$$
\vskip2mm

\noindent Its derivative,  $J^{\prime}$ is of the form
\begin{equation}\label{K}
J^{\prime}(z)= -a + qz +\int_{\mathbb{R}}y(\mathbf{1}_{(-1, 1)}(y)-e^{-zy}) \ \nu(dy),\qquad z\in\mathbb{R}.
\end{equation}
It is clear that $\mid J^{\prime}(0)\mid<+\infty$ if and only if
$$
{\text {(B0)}} \qquad\qquad \int_{\mid y\mid>1}\mid
y\mid\nu(dy)<+\infty,
$$
and $\mid J^{\prime}(z)\mid<+\infty, z>0$  iff
$$
\int_{- \infty}^{-1}|y|e^{z|y|} \nu(dy)<+\infty.
$$
In particular, if the support of the L\'evy measure is bounded from
below then $J^{\prime}$ is well defined and continuous on
$[0,+\infty)$ if $(B0)$ is satisfied. $J^{\prime}$ is automatically
increasing on its domain and its derivative is equal to:
\begin{equation}\label{K'}
J^{\prime\prime}(z)= q +\int_{\mathbb{R}}y^{2}e^{-zy} \nu(dy),\qquad
z\in\mathbb{R}.
\end{equation}
The results on the function $J^{\prime}$ formulated below are explained in detail
in Section \ref{Laplace exponent} in Appendix. Note that the behavior of
$J^{\prime}$ near the origin depends on the behavior of $\nu$ on $[-1,1]^c$.
\begin{prop}\label{Lipschitzowskosc J}
The function $J^{\prime}$ is   Lipschitz   on $[0,z_0]$,   $z_0>0$ if and only if
\begin{gather*}
\text{(L1)} \qquad\qquad \int_{-\infty}^{-1}\mid y\mid^2 e^{z_0\mid y\mid}\nu(dy)<+\infty, \quad \text{and} \quad \int_{1}^{+\infty} y^2\nu(dy)<+\infty.
\end{gather*}
The function $J^{\prime\prime}$ is   Lipschitz  on $[0,z_0]$,   $z_0>0$ if and only if
\begin{gather*}
\text{(L2)} \qquad\qquad \int_{-\infty}^{-1}\mid y\mid^3 e^{z_0\mid y\mid}\nu(dy)<+\infty, \quad \text{and} \quad \int_{1}^{+\infty} y^3\nu(dy)<+\infty.
\end{gather*}
\end{prop}

\begin{prop}\label{bounded}
The function $J^{\prime}$ is bounded on $[0,+\infty)$ iff
\begin{gather*}
(B1)\qquad\qquad \noindent q=0, \quad \text{supp}\{\nu\}\subseteq [0,+\infty) \quad \text{and} \quad  \int_{0}^{+\infty} y \nu (dy)<+ \infty.
\end{gather*}
The function $J^{\prime\prime}$ is bounded on $[0,+\infty)$ iff
\begin{gather*}
(B2)\qquad\qquad\qquad\qquad\qquad \text{supp}\{\nu\}\subseteq [0,+\infty) \quad \text{and} \quad  \int_{1}^{\infty} y^{2} \nu (dy) <\infty.
\end{gather*}
\end{prop}
\vskip2mm

\noindent In the second part of the paper we will need more involved assumptions on the growth of the function $J'$.
\begin{gather}\nonumber
(B3)\qquad\qquad {\rm For}\,{\rm some}\,\, a>0,\,\,b\in\mathbb{R},\,\,\,\,  J^{\prime}(z)\geq a(\ln z)^3+b, \qquad {\rm for}\,{\rm all}\qquad z>0.
\end{gather}
\vskip2mm

\begin{gather}\nonumber
(B4) \qquad\qquad  \limsup_{z\rightarrow\infty} \ \left(\ln z-\bar{\lambda}T^\ast
 J^{\prime}\left(z\right)\right)=+\infty,\qquad 0<T^\ast<+\infty;
\end{gather}
\vskip2mm
\noindent

\noindent If $J^{\prime}$ is a bounded function then (B4) obviously holds. Thus, in particular, (B4) is satisfied for subordinators (increasing L\'evy processes) with possible drifts, see Proposition  4.1 in \cite{BarZab}. However, (B4) does not imply that $J'$ is bounded, see Example 4.2 in \cite{BarZab}. Moreover, we have the following result, see \cite{BarZab}.
\vskip2mm

\begin{prop}\label{tw o subordynatorze}
If $q>0$ or $\nu\{(-b,0)\}>0$, $b>0$ in the
representation \eqref{Laplace transform}, then $J^{\prime}$ satisfies (B3).
\end{prop}

\noindent
This means that each L\`evy process with non-degenerate Wiener part or negative jumps automatically
satisfies (B3). Moreover, if $L$ does not have the Wiener part nor negative jumps then (B4) is affected only
by the behavior of $\nu$ close to zero. To see this, note that
\begin{gather*}
\sup_{z\geq0}\int_{1}^{+\infty}ye^{-zy}\nu(dy)<+\infty,
\end{gather*}
which means that the part of $J^{\prime}$
corresponding to jumps greater than $1$ is bounded.
Thus (B4) in fact depends on the growth of the function
\begin{gather*}
z \ \rightarrow \ \int_{0}^{1}ye^{-zy}\nu(dy)<+\infty.
\end{gather*}

\noindent Below we formulate the conditions (B3) and (B4) explicitly in terms of the measure $\nu$, for the proofs we refer to \cite{BarZab}.
Let us recall that a positive function $M$ {\it varies slowly at $0$} if for any fixed $x>0$
\begin{gather*}
\frac{M(tx)}{M(t)}\longrightarrow 1, \qquad \text{as} \ t\longrightarrow0.
\end{gather*}
If
\begin{gather*}
\frac{f(x)}{g(x)}\longrightarrow 1, \qquad \text{as} \ x\longrightarrow0,
\end{gather*}
we write $f(x)\sim g(x)$.\vskip2mm
\begin{prop}\label{tw glowne Tauber}
Assume that for some $\rho\in(0,+\infty)$,
\begin{enumerate}[] 
\item (B5)\qquad \qquad $\int_{0}^{x}y^2\nu(dy)\sim  x^{\rho} \cdot M(x), \qquad as \ x\rightarrow0, \qquad$
\end{enumerate}
where $M$ is a slowly varying function at $0$.
\begin{enumerate}[i)]
\item If $\rho>1$ then (B4) holds.
\item If $\rho<1$, then (B3) holds.
\item If $\rho=1$, the measure $\nu$ has a density and
\begin{equation}\label{warunek na L}
M(x) \longrightarrow 0 \quad \text{as} \ x\rightarrow 0, \quad and \quad \int_{0}^{1}\frac{M(x)}{x}\ dx=+\infty,
\end{equation}
then (B4) holds.
\end{enumerate}
\end{prop}

\newpage

\part{{HJMM equation with general diffusion}}\label{General case}
By classical results, see e.g. \cite{PeszatZabczyk}, existence of
weak solution to \eqref{general equation} is equivalent to the
existence of a solution to the integral version of \eqref{general
equation}:
\begin{align}\label{semigroup equation}\nonumber
r(t,x)=S_t\left(r_0\right)(x)&+\int_{0}^{t}S_{t-s}\Big(F(r(s))\Big)(x)ds\\[2ex]
&+\int_{0}^{t}S_{t-s}\Big(G(r(s-))\Big)(x)dL(s),\,\,\,\,\,x\geq
0,\,\,\,\,t\in (0,T^\ast],
\end{align}
called mild solution. In (\ref{semigroup equation}),\,  $\{S_t, t\geq 0\}$, stands for the shift semigroup
\begin{gather*}
 S_t(h)(x):=h(t+x),\,\,\,t\geq 0,\,\,x\geq 0,\,\,\,h\in H,
\end{gather*}
acting on the Hilbert space $H$. The equation \eqref{semigroup equation} will be treated here within the
standard SPDE framework for which the crucial role is played by the Lipschitz properties of the transformations $F$ and $G$. Existence of positive
solutions is deduced from abstract results presented in Section \ref{Abstract results}. Theorem \ref{tw o istnieniu przy lokalnych Lipschitzach}
generalizes standard results on existence, see \cite{PeszatZabczyk}, to the case when coefficients have linear growth and are locally Lipschitz. To
obtain positivity of solutions we use Theorem \ref{tw Milian generalized} which is a generalized version of the result of Milian, see \cite{Milian}, and
provides if and only if conditions for positivity in the framework of locally Lipschitz coefficients. As a corollary, in Theorem \ref{tw o warunkach na
positivity} we obtain direct conditions for positivity in our model.

The results on existence of local solutions in $L^{2,\gamma}_{+}$ and
$H^{1,\gamma}_{+}$ are formulated as Theorem \ref{tw o istnieniu rozwiazan lokalnych w L^2} and Theorem \ref{tw o istnieniu rozwiazan lokalnych w H^1}, respectively. They require some regularity properties of the function $g$ as well as local Lipschitz property for $J^{\prime}$ and $J^{\prime\prime}$ which in turn reduce to the integrability conditions (L1), (L2) for the L\'evy measure on the complement of $[-1,1]$. Theorem \ref{rem J^prime locally Lipschitz}, and Theorem \ref{rem J^prime and J^primeprime locally lipschitz}, which are reformulations of the above results, show that, in particular, local solutions exist for the noise with small jumps only and the Wiener process.

For the results on global solutions, which  are formulated as Theorem \ref{tw o istnieniu rozwiazan globalnych w L^2} and Theorem \ref{tw o istnieniu globalnych rozwiazan w H^1}, we need more assumptions. For the space $L^{2,\gamma}_{+}$ boundedness of $J^{\prime}$ on $[0,+\infty)$ is required and for $H^{1,\gamma}_{+}$ boundedness of both $J^{\prime}$ and $J^{\prime\prime}$ is needed. These conditions are rather restrictive and exclude all L\'evy processes which have Wiener part or negative jumps, see Theorem \ref{rem o globalnym istnieniu w L2} and Theorem \ref{rem o globalnym istnieniu w H1}.

Proofs are postponed to Section \ref{Abstract results}.

\section{Formulation of the main  results}\label{Results -
general case}

{We start from a general result on positivity of the solutions to the equation (\ref{semigroup equation}) which  throws some light on the conditions
imposed in the sequel. It is a consequence  of our generalization of an abstract result on positivity due to Milian, see Theorem \ref{tw Milian
generalized}.}

\begin{tw}\label{tw o warunkach na positivity}
Assume that $G$ and $F$ in \eqref{semigroup equation} are
locally Lipschitz in $H$.
Then \eqref{semigroup equation} is positivity preserving if and only if
 \begin{align}\label{positivity 1}
r+g(x,r)u&\ge0\quad\ &&\text{for all}\ r\ge0,\ x\ge0,\ u\in
\mathop{\rm supp} \nu,
\\\nonumber
g(x,0)&=0 \quad\ &&\text{for all}\ x\ge0.
\end{align}
\end{tw}
\vskip4mm

\noindent
\centerline{\underline{LOCAL \ EXISTENCE}} \vskip2mm

\noindent For solvability of the HJMM equation in $L^{2,\gamma}_{+}$ we will need the following conditions on $g$:

\vskip2mm
\[ (G1) \quad
\begin{cases}
  \quad (i) \quad \text{The function g is continuous on} \ \mathbb{R}^2_{+} \ \text{and} \\
\qquad\qquad g(x,0)=0, \ g(x,y)\geq 0, \quad x,y\geq0.\\[2ex]
   \quad (ii) \quad \text{For all} \ x,y\geq 0 \ \text{and} \ u\in
\mathop{\rm supp} \nu:\\
\qquad\qquad   x+g(x,y)u\geq 0.\\[2ex]
 \quad (iii) \quad \text{There exists a constant} \ C>0 \ \text{such that} \\
  \qquad\qquad \mid g(x,u)-g(x,v)\mid\leq C\mid u-v\mid, \qquad  x,u,v\geq0.
 \end{cases}
\]
\vskip2mm

\begin{tw}\label{tw o istnieniu rozwiazan lokalnych w L^2}
 Assume that $J^{\prime}$ satisfies Lipschitz condition in some interval $[0,z_0], z_0>0$ and that $(G1)$ holds.
 Then for arbitrary initial condition $r_0\in L^{2,\gamma}_{+}$ there exists a unique local solution of \eqref{semigroup equation}
 in $L^{2,\gamma}_{+}$.
\end{tw}

{In view of Proposition \ref{Lipschitzowskosc J} we get more explicit result.}

\begin{tw}\label{rem J^prime locally Lipschitz}
 Assume that $(G1)$ holds and either $L$ is a Wiener process or for some $z_0>0$:
\begin{gather*}
\int_{-\infty}^{-1} \mid y\mid^2 e^{z_0\mid y\mid}\nu(dy)<+\infty, \quad \text{and} \quad
\int_{1}^{+\infty}y^2\nu(dy)<+\infty.
\end{gather*}
Then for arbitrary initial condition $r_0\in L^{2,\gamma}_{+}$ there exists a unique local solution of \eqref{semigroup equation}
 in $L^{2,\gamma}_{+}$.
\end{tw}

\vskip4mm
\noindent For local existence in $H^{1,\gamma}_{+}$ we will need more stringent conditions on $g$:

\vskip2mm
\[ (G2) \quad
\begin{cases}
\quad (i) \quad \text{The functions} \ g^{\prime}_x, g^{\prime}_y \
\text{are continuous on} \ \mathbb{R}^2_{+} \
 \text{and}\\
\qquad\qquad  g^{\prime}_x(x,0)=0, \qquad x\geq0.\\[2ex]
\quad (ii) \quad \sup_{x,y\geq 0}\mid g^\prime_y(x,y)\mid<+\infty,\\[2ex]
\quad (iii) \quad \text{There exists a constant} \ C>0 \ \text{such that}\\
\qquad\qquad  \mid g^\prime_x(x,u)-g^\prime_x(x,v)\mid+\mid
g^\prime_y(x,u)-g^\prime_y(x,v)\mid\leq C\mid u-v\mid, \qquad
x,u,v\geq0.
 \end{cases}
\]
\vskip2mm

\begin{tw}\label{tw o istnieniu rozwiazan lokalnych w H^1}
 Assume that $J^{\prime}$ and $J^{\prime\prime}$ satisfy Lipschitz condition in some interval $[0,z_0], z_0>0$ and that $(G1)$ and $(G2)$ hold. Then for arbitrary initial condition $r_0\in H^{1,\gamma}_{+}$ there exists a unique local solution of \eqref{semigroup equation}
 in $H^{1,\gamma}_{+}$.
\end{tw}

{From Proposition \ref{Lipschitzowskosc J} and Proposition \ref{bounded} one can deduce more explicit result.}

\begin{tw}\label{rem J^prime and J^primeprime locally lipschitz}
Assume that $(G1)$ and $(G2)$ hold and for some $z_0>0$
\begin{gather*}
\int_{-\infty}^{-1} \mid y\mid^3 e^{z_0\mid y\mid}\nu(dy)<+\infty, \quad \text{and} \quad \int_{1}^{+\infty}y^3\nu(dy)<+\infty.
\end{gather*}
Then for arbitrary initial condition $r_0\in H^{1,\gamma}_{+}$ there exists a unique local solution of \eqref{semigroup equation}
 in $H^{1,\gamma}_{+}$.
\end{tw}
\vskip4mm

\noindent {Some comments on the imposed conditions are in place now.} \vskip2mm

\noindent If $ supp\{\nu\}\subseteq[0,+\infty)$, that is when $L$ has positive jumps only,  and  $(G1)(i)$ holds then the crucial
positivity condition $(G1)(ii)$ is satisfied. More general result is true.
\vskip2mm

\begin{prop} If for some $m\geq 0$,  $ supp\{\nu\}\subseteq[-m,+\infty)$ and
 $(G1)(i)$ holds then the condition $(G1)(ii)$ holds if and only if
\begin{gather*}
 0\leq g(x,y)\leq \frac{y}{m}, \qquad x,y\geq0.
\end{gather*}
If $\bar{g}(y):=\sup_{x\geq0}g(x,y)<+\infty$, then $(G1)(ii)$ holds if and only if
\begin{gather*}
supp\{\nu\}\subseteq \left[-\inf_{y\geq0}\frac{y}{\bar{g}(y)},+\infty\right).
\end{gather*}
\end{prop}
{\bf Proof:} Indeed, we have $x+g(x,y)\geq x-g(x,y)m\geq0$.\\
Moreover$(G1)(ii)$ holds iff for all $u\in supp\{\nu\}$
\begin{gather}\label{warunek na nosnik}
 u\geq-\inf_{x,y\geq 0}\frac{y}{g(x,y)}=-\inf_{y\geq 0}\frac{y}{\bar{g}(y)}.
\end{gather}
\hfill$\square$

\centerline{\underline{GLOBAL \ EXISTENCE}} \vskip2mm

\noindent We pass now to the global existence results first in  $L^{2,\gamma}_{+}$ and then in $H^{1,\gamma}_{+}$.

\begin{tw}\label{tw o istnieniu rozwiazan globalnych w L^2}
 Assume that $J^{\prime}$ is Lipschitz on some $[0,z_0], z_0>0$ and bounded on $[0,+\infty)$ and that $(G1)$ holds.
 Then for arbitrary $r_0\in L^{2,\gamma}_{+}$ the equation \eqref{semigroup equation} has unique global solution in $L^{2,\gamma}_{+}$.
\end{tw}

{In virtue of Proposition \ref{Lipschitzowskosc J} and Proposition \ref{bounded} we get more explicit result.}

\begin{tw}\label{rem o globalnym istnieniu w L2}
Assume that  $(G1)$ holds and in addition:
\begin{gather*}
q=0, \quad supp\{\nu\}\subseteq [0,+\infty), \quad \int_{0}^{+\infty}\max\{y,y^2\}\nu(dy)<+\infty.
\end{gather*}
Then for arbitrary $r_0\in L^{2,\gamma}_{+}$ the equation \eqref{semigroup equation} has unique global solution in $L^{2,\gamma}_{+}$.
\end{tw}

For global existence in $H^{1,\gamma}_{+}$ we need additional conditions on $g$:

\vskip2mm
\[ (G3) \quad
\begin{cases}
  \quad (i) \quad
\text{Partial derivatives} \ g^{\prime}_y, g^{\prime\prime}_{xy},
g^{\prime\prime}_{yy} \ \text{are bounded on} \ \mathbb{R}^2_{+}.\\[2ex]
   \quad (ii) \quad 0\leq g(x,y)\leq c\sqrt{y}, \qquad x,y\geq0,\\[2ex]
 \quad (iii) \quad \mid g^{\prime}_{x}(x,y)\mid\leq h(x), \qquad x,y\geq 0, \ \text{for some} \ h\in L^{2,\gamma}_+.
 \end{cases}
\]
\vskip2mm

\begin{tw}\label{tw o istnieniu globalnych rozwiazan w H^1}
Let $J^{\prime}, J^{\prime\prime}$ be Lipschitz on some $[0,z_0], z_0>0$ and bounded on $[0,+\infty)$.
Assume that conditions $(G1)$, $(G2)$ and $(G3)$ are satisfied. Then for arbitrary $r_0\in H^{1,\gamma}_{+}$ there exists a unique global solution of \eqref{semigroup equation}
 in $H^{1,\gamma}_{+}$.
\end{tw}

{In virtue of Proposition \ref{Lipschitzowskosc J} and Proposition \ref{bounded} we get more explicit result.}

\begin{tw}\label{rem o globalnym istnieniu w H1}
Assume that conditions $(G1)$, $(G2)$ and $(G3)$ are satisfied and
\begin{gather*}
q=0,\quad supp\{\nu\}\subseteq [0,+\infty), \quad
\int_{0}^{+\infty}\max\{y,y^3\}\nu(dy)<+\infty.
\end{gather*}
Then for arbitrary $r_0\in H^{1,\gamma}_{+}$ there exists a unique global solution of \eqref{semigroup equation}
 in $H^{1,\gamma}_{+}$.
\end{tw}


\section{Proofs of the results}\label{Abstract results}

{The proofs will be based on general  existence and positivity results for  evolution equations:
\begin{gather}\label{SPDE postac ogolna}
  dX=(AX+F(X))dt+G(X)dL,
\end{gather}
with one dimensional L\'evy process $L$ and general transformations $F$, $G$ acting on the Hilbert state space $H$. They are  some improvements of the
classical results.  {Their proofs are given at the end of the present section}. \vskip2mm

\begin{tw}\label{tw o istnieniu przy lokalnych Lipschitzach}
Assume that
\begin{gather*} \|F(x)\|_H+ \|G(x)\|_{H}\le c(1+|x|)
\end{gather*}
for some $c>0$ and for each $R>0$ there exists $c_R>0$ such that for
all $x,y\in H$ satisfying $|x|\le R$, $|y|\le R$,
\begin{gather*}
\|F(x)-F(y)\|+\|G(x)-G(y)\|_{H}\le c_R|x-y|.
\end{gather*}
Then there exists a unique c\`adl\`ag weak solution to the equation \eqref{SPDE postac ogolna}.
\end{tw}

\noindent {The following theorem is an extension of a result of Milian \cite{Milian}   to the equations with locally Lipschitz
coefficients.}

\begin{tw}\label{tw Milian generalized}  Assume that the equation
\eqref{SPDE postac ogolna}, with a Wiener process $L$, admits a solution $X$.
  Assume, in addition, that $A$ generates a strongly continuous semigroup $S_t, t\geq0$ in $H = L^{2}(E, \mu)$, with $\mu$ being a $\sigma$- finite measure on $E$, and that the semigroup preserves positivity.
Assume that for each $R$ there exists a constant $C_R$ such that
  \begin{gather}\label{locally Lipschitz cond. Milian th.}
   \|F(x)-F(y)\|_{H}+\|G(x)-G(y)\|_{H}\leq C_R\|x-y\|_{H}, \quad x,y\in B_R,
  \end{gather}
where $B_R:=\{z\in H; \|z\|_H\leq R\}$. If for each $f\in H_{+}\cap
C_{c}^{\infty}(E)$ and $\varphi\in H_{+}\cap C(E)$ such that
$\langle \varphi,f\rangle=0$ the following holds
\begin{gather}\label{Milian condition 1a}
 \langle F(\varphi),f\rangle\geq 0\\ \label{Milian condition 1b}
\langle G(\varphi),f\rangle=0,
\end{gather}
then $X\geq0$. {Conversely, if all solutions to (\ref{SPDE postac ogolna}), starting from non-negative initial conditions, stay non-negative, then \eqref{Milian condition
1a} and \eqref{Milian condition 1b} hold}.
\end{tw}


\subsection{Proof of Theorem \ref{tw o warunkach na positivity}}
We will use Theorem \ref{tw Milian generalized} {in a similar way as in  \cite{PeszatZabczyk}}.  Let us consider the L\'evy-It\^o
decomposition of $L$
\begin{align*}
L(t)&=at+qW(t)+L_0(t)+L_1(t), \ \text{where}\\
L_0(t)&:=\int_{0}^{t}\int_{\mid y\mid\leq 1}y\hat{\pi}(ds,dy), \ L_1(t):=\int_{0}^{t}\int_{\mid y\mid>1}y\pi(ds,dy),
\end{align*}
and a sequence of its approximations of the form
\begin{gather*}
L^{n}(t)=at+qW(t)-tm_n+(L_0^n(t)+L_1(t)),
\end{gather*}
with $L_0^n(t):=\int_{0}^{t}\int_{\{\frac{1}{n}<\mid y\mid\leq 1\}}y\pi(dy)$ and $m_n:=E[L_0^n(1)]$. Here $\pi$ stands for the random Poisson measure of $L$ and $\hat{\pi}$ for its compensated measure.

The equation \eqref{semigroup equation} preserves positivity if and only if for each $n$ the equation
\begin{align}\label{aproximating equation}\nonumber
dr_n(t,x)=\Big(\frac{d}{dx}r_n(t,x)&+\Big(J^{\prime}\Big(\int_{0}^{x}g(y,r_n(t,y))dy\Big)+a-m_n\Big)g(x,r_n(t,x))\Big)dt\\[1ex]
&+g(x,r_n(t-,x))(dL_0^n(t)+dL_1(t)+qdW(t)),
\end{align}
does. As the sum $L_0^n(t)+L_1(t)$ is a compound Poisson process with jumps greater than $\frac{1}{n}$, the driving noise in \eqref{aproximating equation}
between the jumps is the Wiener process only.  Thus we may use the result of Milian. The conditions
\begin{align*}
&\int_{0}^{+\infty}\Big(J^{\prime}\Big(\int_{0}^{x}g(v,\varphi(v))dv\Big)+a-m_n\Big)g(x,\varphi(x))f(x)e^{\gamma x}dx\geq 0,\\[1ex]
&\int_{0}^{+\infty}g(x,\varphi(x))f(x)e^{\gamma x}dx=0,
\end{align*}
are satisfied for any $\varphi,f\in L^{2,\gamma}$ such that $<\varphi,f>=0$ if and only if $g(x,0)=0$. The solution remains positive in the moment of
jump of $L^n$ if and only if
\begin{gather*}
r+g(x,r)u\geq 0, \qquad r\geq0, u\in \text{supp}\{\nu\}\cup\Big[\frac{1}{n},+\infty\Big)
\end{gather*}
Passing to the limit $n\rightarrow +\infty$ we obtain \eqref{positivity 1}. \hfill$\square$

\subsection{Proofs of Theorem \ref{tw o istnieniu rozwiazan lokalnych w L^2}, Theorem  \ref{tw o istnieniu rozwiazan globalnych w L^2} and Theorem \ref{tw o istnieniu rozwiazan lokalnych w H^1}, Theorem \ref{tw o istnieniu globalnych rozwiazan w H^1}}

For the proofs of the existence results from Section \ref{Results - general case} it is enough to  establish local Lipschitz property and linear growth
for $F,G$ in $L^{2,\gamma}$ and $H^{1,\gamma}$ respectively, formulated as  Proposition \ref{prop local properties L^2} and Proposition \ref{prop local
properties H^1}. Then Theorem \ref{tw o istnieniu rozwiazan lokalnych w L^2} and Theorem \ref{tw o istnieniu rozwiazan lokalnych w H^1} follow from
Theorem \ref{tw o warunkach na positivity} and the fact that locally Lipschitz coefficients imply existence of local solutions, see \cite{PeszatZabczyk}.
Theorem \ref{tw o istnieniu rozwiazan globalnych w L^2} and Theorem \ref{tw o istnieniu globalnych rozwiazan w H^1} can be deduced from Theorem \ref{tw o
istnieniu przy lokalnych Lipschitzach} and Theorem \ref{tw o warunkach na positivity}.\vskip2mm

\subsubsection{Local Lipschitzianity and linear growth of the coefficients in $L^{2,\gamma}$}

\noindent As in the space $L^{2,\gamma}$ the Lipschitz condition of $g$ implies linear growth and Lipschitz property of $G$, below we formulate the
results for $F$ only.
\begin{prop}\label{prop local properties L^2}
Assume that $(G1)$ is satisfied.
\begin{enumerate}[a)]
\item If $J^{\prime}$ is bounded on $[0,+\infty)$
then $F$ has linear growth.
\item If $J^{\prime}$ is
locally Lipschitz then $F$ is locally Lipschitz.
\end{enumerate}
\end{prop}
{\bf Proof of Proposition \ref{prop local properties L^2}:} Let
$C_1:=\sup_{z\geq0}J^{\prime}(z)<+\infty$. \\
\noindent$a)$ The following estimations hold
\begin{align*}
\parallel F(r)\parallel_{L^{2,\gamma}}&=\int_{0}^{+\infty}\left[J^{\prime}\left(\int_{0}^{x}g(y,r(y))dy\right)g(x,r(x))\right]^2e^{\gamma
x}dx\leq C_1\int_{0}^{+\infty}\left[g(x,r(x))\right]^2e^{\gamma x}dx\\[1ex]
&\leq C_1\int_{0}^{+\infty}\left[g(x,r(x))-g(x,0)\right]^2e^{\gamma
x}dx\leq C_1 C^2\int_{0}^{+\infty}r^2(x)e^{\gamma x}dx\leq C_1
C^2\parallel r\parallel^2_{L^{2,\gamma}}.
\end{align*}
$b)$ For any $r,\bar{r}\in L^{2,\gamma}$ we have
\begin{align*}
\parallel F(r)-F(\bar{r})\parallel^2_{L^{2,\gamma}}&=\int_{0}^{+\infty}\left[J^{\prime}\left(\int_{0}^{x}g(y,r(y))dy\right)g(x,r(x))-
J^{\prime}\left(\int_{0}^{x}g(y,\bar{r}(y))dy\right)g(x,\bar{r}(x))\right]^2e^{\gamma
x}dx\\[1ex]
&\leq 2 I_1+2I_2,
\end{align*}
where
\begin{align*}
I_1&:=\int_{0}^{+\infty}\left[J^{\prime}\left(\int_{0}^{x}g(y,r(y))dy\right)-
J^{\prime}\left(\int_{0}^{x}g(y,\bar{r}(y))dy\right)\right]^2g^2(x,r(x))e^{\gamma
x}dx,\\[1ex]
I_2&:=\int_{0}^{+\infty}\left[J^{\prime}\left(\int_{0}^{x}g(y,\bar{r}(y))dy\right)\right]^2\Big(g(x,\bar{r}(x))-
g(y,r(x))dx\Big)^2e^{\gamma x}dx.
\end{align*}
Let us notice that in view of \eqref{calkowalnosc rozwiazania} in
Appendix we have
\begin{gather*}
\int_{0}^{x}g(y,r(y))dy=\int_{0}^{x}\Big(g(y,r(y))-g(y,0)\Big)dy\leq
C\int_{0}^{x}r(y)dy\leq \frac{C}{\sqrt{\gamma}}\parallel
r\parallel_{L^{2,\gamma}}.
\end{gather*}
Denoting by $D=D(\parallel r\parallel_{L^{2,\gamma}},\parallel
\bar{r}\parallel_{L^{2,\gamma}})$ the local Lipschitz constant of
$J^{\prime}$  we thus obtain
\begin{align*}
I_1&\leq
D\int_{0}^{+\infty}\left[\int_{0}^{x}\Big(g(y,r(y))-g(y,\bar{r}(y))\Big)dy\right]^2g^2(x,r(x))e^{\gamma
x}dx\\[1ex]
&\leq D \parallel
g(\cdot,r(\cdot))-g(\cdot,\bar{r}(\cdot))\parallel^2_{L^{2,\gamma}}\int_{0}^{+\infty}g^2(x,r(x))e^{\gamma
x}dx\\[1ex]
&\leq D \parallel
g(\cdot,r(\cdot))-g(\cdot,\bar{r}(\cdot))\parallel^2_{L^{2,\gamma}}\cdot\int_{0}^{+\infty}\Big(g(x,r(x))-g(x,0)\Big)^2e^{\gamma
x}dx\\[1ex]
&\leq DC^2\int_{0}^{+\infty}(r(x)-\bar{r}(x))^2e^{\gamma x}dx\cdot
C^2 \int_{0}^{+\infty}r^2(x)e^{\gamma x}dx\\[1ex]
&\leq DC^4\parallel r-\bar{r}\parallel^2_{L^{2,\gamma}}\parallel
r\parallel^2_{L^{2,\gamma}}.
\end{align*}
Similarly, for a local boundary $B$ of $J^{\prime}$ we get
\begin{gather*}
I_2\leq BC^2 \parallel r-\bar{r}\parallel^2_{L^{2,\gamma}},
\end{gather*}
and thus local Lipschitz property follows.\hfill$\square$\vskip4mm

\subsubsection{Local Lipschitzianity and linear growth of the coefficients in $H^{1,\gamma}$} \noindent Let us start from the auxiliary result.

\begin{lem}\label{H-Lemma}
If $r\in H^{1,\g}$ then
$$
\sup_{x\ge0}|r(x)|\le2\Bigl(\frac1\g\Bigr)^{1/2}\|r\|_{H^{1,\g}}\,.
$$
\end{lem}
\noindent {\bf Proof of Lemma \ref{H-Lemma}:} Integrating by parts
gives
$$
\int_0^x y\,\frac{dr(y)}{dy}\,dy=yr(y)\big|_0^x - \int_0^x r(y)\,dy,
$$
and thus
\begin{align*}
|xr(x)| &\le \Bigl|\int_0^x y\,\frac{dr(y)}{dy}\,dy\Bigr|
+\Bigl|\int_0^x r(y)\,dy\Bigr|\\[2ex]
{}&\le \Bigl(\int_0^{+\infty}e^{-\g y}y^2\,dy\Bigr)^{1/2}
\Bigl(\int_0^{+\infty}e^{\g
y}\Bigl(\frac{dr(y)}{dy}\Bigr)^2\,dy\Bigr)^{1/2}+
\Bigl(\int_0^{+\infty}e^{-\g y}\,dy\Bigr)^{1/2}
\Bigl(\int_0^{+\infty}e^{\g y}r^2(y)\,dy\Bigr)^{1/2}\\[2ex]
&\le\Bigl(\frac2{\g^3}\Bigr)^{1/2}\|r\|_{H^{1,\g}}+
\Bigl(\frac1{\g}\Bigr)^{1/2}\|r\|_{L^{2,\g}}.
\end{align*}
In particular
$$
\lim_{x\to+\infty}r(x)=0.
$$
Moreover,
\begin{align*}
|r(x)-r(0)|&=\Bigl|\int_0^x \frac{dr}{dy}(y)\,dy\Bigr| \le \int_0^x
e^{-\g y/2}
e^{\g y/2}\Bigl|\frac{dr}{dy}(y)\Bigr|\,dy\\[2ex]
&\le \Bigl(\int_0^{+\infty}e^{-\g y}\,dy\Bigr)^{1/2}
\Bigl(\int_0^{+\infty}e^{\g
y}\Bigl(\frac{dr}{dy}(y)\Bigr)^2\,dy\Bigr)^{1/2}
\leq\Bigl(\frac1\g\Bigr)^{1/2}\|r\|_{H^{1,\g}}\,.
\end{align*}
Consequently,
$$
\mid r(0)\mid \le \Bigl(\frac1\g\Bigr)^{1/2}\|r\|_{H^{1,\g}},
$$
and therefore
$$
\sup_{x\ge0}|r(x)|\le 2\Bigl(\frac1\g\Bigr)^{1/2}\|r\|_{H^{1,\g}}\,.
\eqno\square
$$

\begin{prop}\label{prop local properties H^1}
Assume that $(G1)$ is satisfied.
\begin{enumerate}[a)]
\item If $J^{\prime}$ and $J^{\prime\prime}$ are bounded on
$[0,+\infty)$ and $(G3)$ holds
then $G$ and $F$ have linear growth.
\item If $J^{\prime}$ and $J^{\prime\prime}$ are
locally Lipschitz and $(G2)$ holds
then $F$ and $G$ are locally Lipschitz.
\end{enumerate}
\end{prop}

\noindent
{\bf Proof of Proposition \ref{prop local properties
H^1}:} $(a)$ Linear growth of $G$ follows from the estimation
\begin{align}\label{oszacowanie linear growth G in H^1}\nonumber
 \int_{0}^{+\infty}\mid\frac{d}{dx}g(x,r(x))\mid^2&e^{\gamma x}dx=\int_{0}^{+\infty}\left[g^{\prime}_x(x,r(x))+g^{\prime}_{y}(x,r(x))r^\prime(x)\right]^2e^{\gamma x}dx \\[1ex]\nonumber
 &\leq 2\int_{0}^{+\infty}[h(x)]^2e^{\gamma x}dx+2\sup_{x,r\geq 0}\mid g^{\prime}_{y}(x,r)\mid^2\int_{0}^{+\infty}\mid r^\prime(x)\mid^2e^{\gamma x}dx\\[1ex]
 &\leq 2\parallel h\parallel^2_{L^{2,\gamma}}+2\sup_{x,r\geq 0}\mid g^{\prime}_{y}(x,r)\mid^2\cdot\parallel r\parallel^2_{H^{1,\gamma}}.
\end{align}
To show linear growth of $F$ let us start with the inequality
\begin{align*}
 \int_{0}^{+\infty}\frac{d}{dx}\Big(J^{\prime}\Big(\int_{0}^{x}g(v,r(v))dv\Big)g(x,r(x))\Big)\leq 2\int_{0}^{+\infty}\Big| J^{\prime\prime}\Big(\int_{0}^{x}g(v,r(v))dv\Big)g^2(x,r(x))\Big|^2 e^{\gamma x}dx\\[1ex]
 +\int_{0}^{+\infty}\Big|J^{\prime}\Big(\int_{0}^{+\infty}g(v,r(v))dv\Big)[g^\prime_x(x,r(x))+g^{\prime}_{y}(x,r(x))r^\prime(x)]\Big|^2e^{\gamma x}dx.
\end{align*}
The second integral can be estimated in the same way as \eqref{oszacowanie linear growth G in H^1}. Linear growth of the first integral follows from the inequality
\begin{gather*}
\int_{0}^{+\infty}\mid g(x,r(x))\mid^4e^{\gamma
x}dx\leq\sup_{x,r\geq0}\Big|\frac{g(x,r)}{\sqrt{r}}\Big|^4\int_{0}^{+\infty}
\mid r(x)\mid^2e^{\gamma x}dx.
\end{gather*}

\noindent
$(b)$
To get the required estimation for $G$ we need to estimate
\begin{gather*}
I_0:=\int_{0}^{+\infty}\left[g^{\prime}_{y}(x,r(x))r^{\prime}(x)-g^{\prime}_{y}(x,\bar{r}(x))\bar{r}^{\prime}(x)\right]^2e^{\gamma
x}dx.
\end{gather*}
Using Lemma \ref{H-Lemma} we obtain the following inequalities
\begin{align*}
I_0&\leq 2
\int_{0}^{+\infty}\left|g^{\prime}_{y}(x,r(x))\right|^2\left|r^{\prime}(x)-\bar{r}^{\prime}(x))\right|^2e^{\gamma
x}dx+\int_{0}^{+\infty}\left|g^{\prime}_{y}(x,r(x))-g^{\prime}_{y}(x,\bar{r}(x))\right|^2\left|\bar{r}^{\prime}(x))\right|^2e^{\gamma
x}dx\\[1ex]
&\leq 2\sup_{x,r\geq0}\mid
g^{\prime}_{y}(x,r)\mid^2\int_{0}^{+\infty}\mid
r^{\prime}(x)-\bar{r}^{\prime}(x)\mid^2e^{\gamma x}dx\\[1ex]
&+ 2\left(\sup_{x,u,v\geq0}\frac{\mid
g^{\prime}_{y}(x,u)-g^{\prime}_{y}(x,u)\mid}{\mid u-v\mid}\right)^2
\int_{0}^{+\infty}\mid
r(x)-\bar{r}(x)\mid^2(\bar{r}^\prime(x))^2e^{\gamma x}dx\\[1ex]
&\leq 2\sup_{x,r\geq0}\mid
g^{\prime}_{y}(x,r)\mid^2\cdot\parallel r-\bar{r}\parallel^2_{H^{1,\gamma}}\\[1ex]
&+ 2\left(\sup_{x,u,v\geq0}\frac{\mid
g^{\prime}_{y}(x,u)-g^{\prime}_{y}(x,u)\mid}{\mid u-v\mid}\right)^2
\frac{4}{\gamma}\parallel
r-\bar{r}\parallel^2_{H^{1,\gamma}}\cdot\parallel
\bar{r}\parallel^2_{H^{1,\gamma}},
\end{align*}
and thus local Lipschitz property for $G$ follows.

To show the same for $F$ it is sufficient to
show the Lipschitz estimation for the formula
\begin{align*}
I:=\int_{0}^{+\infty}\left[\frac{d}{dx}\left\{\left(J^{\prime}\left(\int_{0}^{x}g(y,r(y))dy\right)g(x,r(x))
-J^{\prime}\left(\int_{0}^{x}g(y,\bar{r}(y))dy\right)g(x,\bar{r}(x))\right)\right\}\right]^2e^{\gamma
x}dx.
\end{align*}
By explicit calculations we obtain
\begin{gather*}
I\leq 3I_1+3I_2+3I_3,
\end{gather*}
where
\begin{align*}
I_1&:=\int_{0}^{+\infty}\left[J^{\prime\prime}\left(\int_{0}^{x}g(y,r(y))dy\right)g^2(x,r(x))-
J^{\prime\prime}\left(\int_{0}^{x}g(y,\bar{r}(y))dy\right)g^2(x,\bar{r}(x))\right]^2e^{\gamma x}dx,\\[1ex]
I_2&:=\int_{0}^{+\infty}\left[J^{\prime}\left(\int_{0}^{x}g(y,r(y))dy\right)g^{\prime}_x(x,r(x))-
J^{\prime}\left(\int_{0}^{x}g(y,\bar{r}(y))dy\right)g^{\prime}_x(x,\bar{r}(x))\right]^2e^{\gamma x}dx,\\[1ex]
I_3&:=\int_{0}^{+\infty}\left[J^{\prime}\left(\int_{0}^{x}g(y,r(y))dy\right)g^{\prime}_{y}(x,r(x))\cdot
r^{\prime}(x)-
J^{\prime}\left(\int_{0}^{x}g(y,\bar{r}(y))dy\right)g^{\prime}_{y}(x,\bar{r}(x))\cdot\bar{r}^{\prime}(x)\right]^2e^{\gamma
x}dx.
\end{align*}
We can estimate $I_1$ in the same way as
in the proof of Proposition \ref{prop local properties L^2}.
With the use of $(i)$ and $(ii)$ one obtains the estimate for $I_2$. $I_3$
can also be estimated in the same way as in $L^{2,\gamma}$ provided
that we have additional inequalities for
\begin{gather*}
\int_{0}^{+\infty}\left[g^{\prime}_{y}(x,r(x))r^{\prime}(x)-g^{\prime}_{y}(x,\bar{r}(x))\bar{r}^{\prime}(x)\right]^2e^{\gamma
x}dx
\end{gather*}
which is exactly $I_0$ and is estimated above, and
\begin{gather*}
I_4:=\int_{0}^{+\infty}\left[g^{\prime}_{y}(x,\bar{r}(x))\bar{r}^{\prime}(x)\right]^2e^{\gamma
x}dx.
\end{gather*}
{Estimation for $I_4$ follows from   the bound on
$g^{\prime}_{y}$}. \hfill$\square$

\subsection{Proof of Theorem \ref{tw o istnieniu przy lokalnych Lipschitzach}}\label{Proofs of the abstract results}

Let $F_n, G_n,\, n=1,2,\ldots$ be such that
\begin{enumerate}[(i)]
\item $F_n(x)=F(x)$ and $G_n(x)= G(x)$ if $|x|\le n$,
\item for all $t>0$ and $x\in H$,
$$
\|F_n(x)\|+\|G_n(x)\| \le c(1+|x|),
$$
\item there is a constant $c_n$ such that for all $x,y\in H$,
$$
\|F_n(x)-F_n(y)\|+\|(G_n(x)-G_n(y))\| \le c_n|x-y|.
$$
\end{enumerate}
By Theorem 9.29 of \cite{PeszatZabczyk}, the equation obtained from \eqref{SPDE postac ogolna} by replacing $F$ and $G$ by $F_n$ and $G_n$, has a unique
c\`adl\`ag solution $X_n$ starting from any $x_0\in H$ and satisfying the estimation
\begin{gather}\label{oszacowanie drugiego momentu}
\sup_{t\le T} \mathbf{E} \|X_n(t)\|^2 \le C\left(1+\|x_0\|^2\right),
\end{gather}
for some $C>0$.

Let
$$
\tau_n:= \inf\{t\le t\colon \|X_n(t)\|>n\}.
$$
On the time interval $[0,\tau_n)$, the trajectories of $X_n$ are contained in the ball $B(0,n)$ in $H$ with center at $0$ and radius $n$, and therefore
$X_n$ satisfies $\eqref{SPDE postac ogolna}$. In particular, for all $m>n$, $X_m$ and $X_n$ coincide on $[0,\tau_n)$. Define $X(t)= X_n(t) $ if
$t<\tau_n$. Note that $X$ is well defined. To finish the proof it is enough to show that
$$
\lim_{n\to \infty}P\left( \sup_{s\le t} \|X_n(t)\|>n\right)=0.
$$
Let $n$ be such that $\|X(0)\|\le n/3$ for $t\le T$. Then
$$
\begin{aligned}
&P\left(\sup_{t\le T} \|X_n(t)\|>n\right) \le P\left( \sup_{t\le T} \left\|
\int_0^t S(t-s) F_n(X_n(s)) ds \right\| >\frac{n}{3}\right) \\
&\qquad + P\left(\sup_{t\le T} \left\| \int_0^t S(t-s) G_n(X_n(s-)) dM(s)\right\| >\frac n3 \right) := I_1+I_2.
\end{aligned}
$$
However, for a constant $\hat{c}$ independent of $n$,
$$
\sup_{t\le T} \left\| \int_0^t S(t-s) F_n(X_n(s)) ds \right\| \le \hat{c}\left(1+ \int_0^T \|X_n(s)\| ds\right),
$$
and hence, by Chebyshev's inequality and \eqref{oszacowanie drugiego momentu}, there is a constant $\hat{\hat{c}}$ such that
$$
\begin{aligned}
I_1 &\le \frac{3\hat{c}}{n} \left (1 + \int_0^T \mathbf{E} \|X_n(s)\| ds\right) \\
& \le \frac{3\hat{c}}{n} \left( 1+\int_0^T \left( \mathbf{E}\|X_n(s)\|^2\right)^{1/2} ds\right) \\
& \le \frac {3\hat{\hat{c}}}n \left( 1 +\|x_0\|^2\right)^{1/2} .
\end{aligned}
$$
Hence $I_1\to 0$ as $n\to \infty$. By Kotelenez's inequality (see e.g. \cite{PeszatZabczyk}) and \eqref{oszacowanie drugiego momentu} there is a constant
$\tilde{c}$ such that
$$
\begin{aligned}
I_2 &\le \left( \frac 3 n\right)^2 \tilde{c} \ \mathbf{E} \int_0^T \| G_n(X_n(s)) \|^2 ds \\
&\le 2c\left(\frac 3n \right)^2\tilde{c} \int_0^T \left(1 + \mathbf{E} \| X_n(s)\|^2 \right) ds\\
&\le \tilde{\tilde{c}}\left(\frac 3 n\right)^2  \left( 1 + \|x_0\|^2\right).
\end{aligned}
$$
Hence $I_2\to 0$ as $n\to \infty$ and the assertion follows. \hfill$\square$

\subsection{Proof of Theorem \ref{tw Milian generalized}}\label{Milian}

\noindent {\bf Proof of Theorem \ref{tw Milian generalized}:} We use the original result of Milian \cite{Milian}. Let us consider a sequence of
transformations
\begin{align*}
 F^n(x):=F(x)h^n(\parallel x\parallel); \qquad G^n(x):=F(x)h^n(\parallel x\parallel),
\end{align*}
where
\[
h^n(z)= \
\begin{cases}
1 \quad &\text{for} \ z\in[0,n), \\
2-\frac{z}{n} \quad &\text{for} \ z\in[n,2n),\\
0 \quad &\text{for} \ z\geq 2n.
\end{cases}
\]
One can check that $h^n$ is Lipschitz for each $n$ and thus $F^n, G^n$ are Lipschitz on $H$. The following hold
\[
\langle F^n(\varphi),f\rangle = \
\begin{cases}
\langle F(\varphi),f\rangle\geq0 \quad &\text{if} \ \parallel\varphi\parallel<n, \\
h^n(\parallel\varphi\parallel)\langle F(\varphi),f\rangle\geq0 \quad &\text{if} \ n\leq\parallel\varphi\parallel<2n,\\
0 \quad &\text{for} \ z\geq 2n,
\end{cases}
\]
and $\langle G^n(\varphi),f\rangle=0$. Therefore it follows that the solution $X^n$ of the equation \eqref{SPDE postac ogolna} with $F,G$ replaced by
$F^n,G^n$ is non-negative. But
\begin{gather*}
X^n=X\mathbf{1}_{B_n}(\parallel X\parallel),
\end{gather*}
which implies that $X$ is non-negative on each ball. Passing to the limit with the radius and using the fact that $X$ is bounded we obtain positivity of
$X$. Using the arguments in the opposite direction we get necessity of \eqref{Milian condition 1a}, \eqref{Milian condition 1b}. \hfill$\square$

\newpage
\part{HJMM equation with linear diffusion}\label{Linear case}

In this part we assume that
$$
g(x,y)=\lambda(x)y, \qquad x,y\geq 0
$$
where $\lambda(\cdot)$ is a continuous function. Then the weak version of \eqref{basic equation} is of  the form:
\begin{align}\label{semigroup equation linear}\nonumber
r(t,x)=S_t(r_0)(x)&+\int_{0}^{t}S_{t-s}\left(J^{\prime}\Big(\int_{0}^{x}\lambda(v)r(s,v)dv\Big)\lambda(x)r(s,x)\right)ds\\[2ex]
&+\int_{0}^{t}S_{t-s}\Big(\lambda(x)r(s-,x)\Big)dL(s),\,\,\,\,\,x\geq
0,\,\,\,\,t\in (0,T^\ast].
\end{align}
\vskip2mm

\noindent The following two conditions (B3) and (B4), already introduced in the Preliminaries, play an essential role in the analysis of  existence of
the global solutions to the equation \eqref{semigroup equation linear}.  Roughly speaking  solutions explode if,\vskip2mm

\noindent
\begin{gather*}
(B3)\quad {\rm For}\,{\rm some}\,\, a>0,\,\,b\in\mathbb{R},\,\,\,\,  J^{\prime}(z)\geq a(\ln z)^3+b, \qquad {\rm for}\,{\rm all} z>0,
\end{gather*}
\vskip2mm

\noindent and global solutions exist if,
\begin{gather*}
(B4)\qquad \limsup_{z\rightarrow\infty} \ \left(\ln z-\bar{\lambda}T^\ast
 J^{\prime}\left(z\right)\right)=+\infty,\qquad 0<T^\ast<+\infty.
\end{gather*}
\vskip2mm

\noindent Results on local existence are formulated as Theorem \ref{local1} and  Theorem \ref{local2} and follow from the general results of the first
part. Theorem \ref{tw o eksplozjach} formulates conditions for non-existence of global solutions and is inspired by a similar result in \cite{BarZab}.
Subsequent results concern global solutions, see Theorem \ref{tw o istnieniu} , strong solutions, see Theorem \ref{tw o silnym rozwiazaniu} as well as
their uniqueness, see Theorem \ref{tw o jedynosci}.\vskip2mm

{Some existence results on global solution to \eqref{semigroup equation linear}  can  be deduced from results of Part~I like Theorem \ref{tw o istnieniu rozwiazan globalnych w
L^2} or   Theorem \ref{tw o istnieniu globalnych rozwiazan w H^1} however under very   restrictive conditions on $J^{\prime}$.  In fact, we have the
following elementary observation.
\begin{prop}\label{Lipschitz}
If the drift transformation $F$ defined by \eqref{volatility and drift}  is of  linear growth   in $L^{2,\gamma}$,  then $J'$ is bounded on $[0,
+\infty)$. In particular
$$
q=0, \quad \text{supp}\{\nu\}\subseteq [0,+\infty) \quad \text{and} \quad  \int_{0}^{+\infty} y \nu (dy)<+ \infty.
$$
\end{prop}
{\bf Proof: } Assume, to the contrary, that $J^{\prime}$ is unbounded and define
\begin{gather*}
 r_n(x)=n\mathbf{1}_{[1,3]}(x), \qquad n=1,2,... \ .
\end{gather*}
As for  sufficiently large $z\geq 0$ the function $(J^{\prime}(z))^2$ is increasing  we have, for large $n$
\begin{gather*}
\frac{\|F(r_n)\|^2}{\|r_n\|^2}=\frac{\int_{1}^{3}\Big( J^{\prime}\Big(\int_{1}^{x}\lambda(y)n dy\Big)\Big)^2\lambda^2(x)n^2e^{\gamma
x}dx}{n^2\int_{1}^{3}e^{\gamma x}dx}\geq\frac{\underline{\lambda}^2\int_{1}^{3}\Big(J^{\prime}\Big(\underline{\lambda} n(x-1)\Big)\Big)^2e^{\gamma
x}dx}{\int_{1}^{3}e^{\gamma x}dx}.
\end{gather*}
Since,
\begin{gather*}
 \int_{1}^{3}\Big(J^{\prime}\Big(\underline{\lambda} n(x-1)\Big)\Big)^2e^{\gamma x}dx\geq \int_{2}^{3}\Big(J^{\prime}\Big(\underline{\lambda}
  n(x-1)\Big)\Big)^2e^{\gamma x}dx\geq \Big(J^{\prime}(\underline{\lambda}n)\Big)^2\int_{2}^{3}e^{\gamma x}dx\underset{n}{\longrightarrow} +\infty,
\end{gather*}
the main claim holds. The rest follows from Proposition \ref{bounded}. \hfill$\square$}

\section{Formulation of the main results}\label{Formulation of the main results}

\centerline{\underline{EXISTENCE OF LOCAL SOLUTIONS}}\label{Existence of local solutions}\vskip2mm

The following theorem is a direct consequence of  Theorem \ref{tw o istnieniu rozwiazan lokalnych w L^2}.
\begin{tw}\label{local1} Assume that:
\begin{enumerate}[]
\item$(\Lambda 0)\qquad \lambda\,\, {\text {is continuous and}} \,\,\inf_{x\geq0}\lambda(x)= \underline{\lambda} >0 , \,\,\,\, \sup_{x\geq0}\lambda(x)=\bar{\lambda} < + \infty,$
\end{enumerate}
\begin{enumerate}[]
\item$(\Lambda 1)\qquad \qquad  \mathop{\rm supp} \nu\subseteq [-\frac{1}{\bar{\lambda}},+\infty)$
\end{enumerate}
\begin{enumerate}[]
\item$(L1)\qquad \qquad    \int_{1}^{+\infty}y^2\nu(dy)<+\infty,$
\end{enumerate}
hold. Then there exists a unique local weak solution to the equation \eqref{semigroup equation linear} taking values in the space $L^{2,\gamma}_{+}$.
\end{tw}
\noindent In the formulation of the theorem  a simplified, but  under $(\Lambda 0), (\Lambda 1)$, equivalent  version of the condition (L1) from the Preliminaries was used.
In fact the positivity assumptions $(G1) (i)$, $(G1) (ii)$ follow from $(\Lambda 0), (\Lambda 1)$  and the assumption $(G1)(iii)$ follows from $(\Lambda 0) $. Local Lipschitzianity is a consequence of $(L1)$, see Proposition \ref{Lipschitzowskosc J} and Proposition \ref{prop local properties L^2}. \vskip2mm

\noindent Similarly as a  consequence of Theorem \ref{tw o istnieniu rozwiazan lokalnych w H^1} we obtain the following local existence result in $H^{1,\gamma}_{+}$.

\begin{tw}\label{local2}
Assume that conditions $(\Lambda 0), (\Lambda 1)$,
\begin{enumerate}[]
\item$(\Lambda 2)\qquad \qquad \lambda ,  \lambda^{\prime}  {\text{are bounded and  continuous on}}\,\,\,  \mathbb{R}_{+},$
\end{enumerate}
and
\begin{enumerate}[]
\item$ (L2) \qquad \qquad   \int_{1}^{+\infty}y^3\nu(dy)<+\infty,$
\end{enumerate}
are satisfied. Then there exists a unique local weak solution to the equation \eqref{semigroup equation linear} taking values in the space
$H^{1,\gamma}_{+}$.
\end{tw}
\vskip4mm

\centerline{\underline{NON-EXISTENCE OF GLOBAL SOLUTIONS IN
$H^{1,\gamma}_{+}$}}\vskip2mm

\noindent Our first result on global solutions is of negative type.

 \begin{tw}\label{tw o eksplozjach}
Assume that conditions $(\Lambda 0), (\Lambda 1)$ ,
\begin{enumerate}[]
\item$(\Lambda 3)\qquad  \lambda ,\lambda^{\prime} ,\lambda^{\prime\prime},\,\, \text{are bounded and continuous on} \ \mathbb{R}_{+},$
\end{enumerate}
 \begin{enumerate}[]
\item $(B0) \qquad  \int_{1}^{+\infty} y\nu(dy)<+\infty ,$
\end{enumerate}
\begin{enumerate}[]
\item $(B3)\quad  J^{\prime}(z)\geq a(\ln z)^3+b, \qquad {\rm for}\,{\rm some}\,\, a>0,\,\,b\in\mathbb{R},\,\,\,\,  {\rm and}\,{\rm all}\,\, z>0$
\end{enumerate}
are satisfied.

\noindent Then,  for some  $k>0$ and all  $r_{0}(\cdot)\in
H^{1,\gamma}_{+}$ such that $r_0(x)\geq k,\, \forall
x\in[0,T^{\ast}]$, the global  solution in $H^{1,\gamma}_{+}$ of
\eqref{semigroup equation linear} does not exist on the interval
$[0, T^{\ast}]$.
\end{tw}
\vskip2mm

\noindent {It follows from  Theorem \ref{local2} and Theorem \ref{tw
o eksplozjach} that if  conditions $(\Lambda 0), (\Lambda 1)$,
$(\Lambda 2)$, $(\Lambda 3)$, (B0), (B3) and (L2) hold then any
local solution in $H^{1,\gamma}_{+}$ explodes.}

\vskip2mm \noindent The theorem  remains true if the condition $(B3)$ is replaced by a stronger but a more explicit condition on the measure $\nu$, see
Proposition \ref{tw glowne Tauber} .
\begin{tw}\label{tw o eksplozjach2}
Assume that conditions $(\Lambda 0), (\Lambda 1)$ ,
\begin{enumerate}[]
\item$(\Lambda 3)\qquad  \lambda ,\lambda^{\prime} ,\lambda^{\prime\prime},\,\, \text{are bounded and continuous on} \ \mathbb{R}_{+},$
\end{enumerate}
 \begin{enumerate}[]
\item $(B0) \qquad  \int_{1}^{+\infty}y\nu(dy)<+\infty ,$
\end{enumerate}
\begin{enumerate}[]
\item $(B5)\qquad \int_{0}^{x}y^2\nu(dy)\sim  x^{\rho} \cdot  M(x), \qquad as \ x\rightarrow0,$
\end{enumerate}
$\text{where}\,\, M \text{is a slowly varying function}  ,$ at $0$ and $\rho <1,$ are satisfied.\vskip2mm

\noindent Then,  for some  $k>0$ and all  $r_{0}(\cdot) \in
H^{1,\gamma}_{+}$  such that $r_0(x)\geq k,\, \forall
x\in[0,T^{\ast}],$ the global  solution in $H^{1,\gamma}_{+}$ of
\eqref{semigroup equation linear} does not exist on the interval
$[0, T^{\ast}]$.
\end{tw}

\vskip4mm

\centerline{\underline{EXISTENCE OF GLOBAL SOLUTIONS}}\vskip2mm

\noindent We have the following existence result in which the key role is played by the logarithmic growth condition (B4).  Condition (B2), which appears
in its formulation, was introduced in Proposition \ref{bounded}.
\begin{tw}\label{tw o istnieniu}
Assume that $(\Lambda 0), (\Lambda 1)$ and conditions
\item$(\Lambda 2)\qquad \qquad \lambda ,  \lambda^{\prime}  {\text{are bounded and  continuous on}}\,\,\,  \mathbb{R}_{+},$
\item $(B0) \qquad\qquad  \int_{1}^{+\infty} y\nu(dy)<+\infty ,$
\item $(B4)\qquad\qquad  \limsup_{z\rightarrow\infty} \ \left(\ln z-\bar{\lambda}T^\ast
 J^{\prime}\left(z\right)\right)=+\infty,\qquad 0<T^\ast<+\infty ,$
\noindent hold.

\noindent(a) If $r_0\in L^{2,\gamma}_{+}$ then there
exists a solution to \eqref{semigroup equation linear} taking values
in the space $L^{2,\gamma}_{+}$.\vskip2mm

\noindent (b) Assume, in addition, that
\item$(\Lambda 3)\qquad  \lambda ,\lambda^{\prime} ,\lambda^{\prime\prime},\,\, \text{are bounded and continuous on} \ \mathbb{R}_{+},$

\item$(B2)\qquad   \text{supp}\{\nu\}\subseteq [0,+\infty) \quad \text{and} \quad  \int_{1}^{\infty} y^{2} \nu (dy) <\infty.$\vskip2mm

\noindent If $r_0\in H^{1,\gamma}_{+}$ then there exists a solution
to \eqref{semigroup equation linear} taking values in the space
$H^{1,\gamma}_{+}$.
\end{tw}
\vskip4mm

\noindent {The condition (B4) implies, like in the general diffusion
case, that the process $L$ should be without Wiener part and without
negative jumps. On the other hand Theorem \ref{tw o eksplozjach}
shows that under (B3) the  absence of the Wiener part and of the
negative jumps in the decomposition of $L$ is also necessary for
existence in the linear case in $H^{1,\gamma}_{+}$.}

The theorem remains true if the condition $(B4)$ is replaced by a
stronger but a specific condition on the measure $\nu$, see
Proposition \ref{tw glowne Tauber} .
\begin{tw}\label{tw o istnieniu2}
Assume that $(\Lambda 0), (\Lambda 1)$ and conditions
\item$(\Lambda 2)\qquad \qquad \lambda ,  \lambda^{\prime}  {\text{are bounded and  continuous on}}\,\,\,  \mathbb{R}_{+},$
\item $(B0) \qquad  \int_{1}^{+\infty} y\nu(dy)<+\infty ,$
\begin{enumerate}[]
\item $(B5)\qquad \int_{0}^{x}y^2\nu(dy)\sim  x^{\rho} \cdot  M(x), \qquad as \ x\rightarrow0,$
\end{enumerate}
$\text{where}\,\, M \text{is a slowly varying function}  ,$ at $0$ and $\rho >1,$ are satisfied.\vskip2mm

\noindent(a) If $r_0\in L^{2,\gamma}_{+}$ then there exists a
solution to \eqref{semigroup equation linear} taking values in the
space $L^{2,\gamma}_{+}$.\vskip2mm

\noindent (b) Assume, in addition, that
\item$(\Lambda 3)\qquad  \lambda ,\lambda^{\prime} ,\lambda^{\prime\prime},\,\, \text{are bounded and continuous on} \ \mathbb{R}_{+},$

\item$(B2)\qquad   \text{supp}\{\nu\}\subseteq [0,+\infty) \quad \text{and} \quad  \int_{1}^{\infty} y^{2} \nu (dy) <\infty.$\vskip2mm

\noindent If $r_0\in H^{1,\gamma}_{+}$ then there exists a solution
to \eqref{semigroup equation linear} taking values in the space
$H^{1,\gamma}_{+}$.
\end{tw}
\vskip3mm

Our assumptions implying global existence are not very restrictive.
The condition $(B4)$ is weaker than the requirement  that
$J^{\prime}$ is bounded, which was necessary for the standard
contraction principle approach, see Proposition \ref{Lipschitz}.
Moreover, the assumptions  do not  imply local Lipschitz property of
the coefficients. In Theorem \ref{tw o istnieniu} we need (B4) and
integrability of $\nu$ outside of the unit ball, that is
\begin{gather}\label{zalozenia o istnieniu L^2}
\int_{1}^{+\infty}y\nu(dy)<+\infty,
\end{gather}
for the space $L^{2,\gamma}_{+}$ and
\begin{gather}\label{zalozenia o istnieniu H^1}
supp\{\nu\}\subseteq [0,+\infty) \quad \text{and} \quad \int_{1}^{+\infty}y^2\nu(dy)<+\infty,
\end{gather}
for $H^{1,\gamma}_{+}$. It is clear that
\begin{gather}\label{pierwsza nieimplikacja}
 \eqref{zalozenia o istnieniu L^2}\quad \nRightarrow\quad  \int_{1}^{+\infty}y^2\nu(dy)<+\infty,
\end{gather}
and
\begin{gather}\label{druga nieimplikacja}
 \eqref{zalozenia o istnieniu H^1}\quad \nRightarrow\quad  \int_{1}^{+\infty}y^3\nu(dy)<+\infty.
\end{gather}
Under the condition $supp\{\nu\}\subseteq [-\frac{1}{\bar{\lambda}},+\infty)$, the right hand sides of \eqref{pierwsza nieimplikacja}, \eqref{druga
nieimplikacja} are equivalent to conditions (L1), (L2) which in turn correspond to local Lipschitz properties of $F$ and $G$ in $L^{2,\gamma}$, resp.
$H^{1,\gamma}$. On the other hand, as explained in the Preliminaries, (B4) is related to the behavior of $\nu$ close to zero. Thus for each L\'evy process
satisfying (B4) and \eqref{zalozenia o istnieniu L^2} (or \eqref{zalozenia o istnieniu H^1}) there exists a global solution in $L^{2,\gamma}_{+}$, (resp.
$H^{1,\gamma}_{+}) $ but $F,G$ are not locally Lipschitz. \vskip2mm

\centerline{\underline{EXISTENCE OF STRONG SOLUTIONS IN
$H^{1,\gamma}_{+}$}}\vskip2mm

\noindent Under additional conditions we can establish existence of the strong solutions to \eqref{basic equation}.
\begin{tw}\label{tw o silnym rozwiazaniu}
Assume that $\lambda(x)\equiv\lambda$ is constant and all
assumptions of Theorem \ref{tw o istnieniu} (b) are satisfied.
 Then the weak non-exploding solution given by Theorem \ref{tw o istnieniu} (b) is a strong solution of \eqref{basic equation}.
\end{tw}
\vskip4mm

\centerline{\underline{UNIQUENESS OF THE GLOBAL SOLUTION IN
$H^{1,\gamma}_{+}$}} \vskip2mm

\noindent Assumptions of Theorem \ref{tw o istnieniu} do not imply, in general, the uniqueness of the solutions. Also this property  does not follow from the uniqueness of the local solutions. Thus the following theorem  cannot be deduced from the contraction principle.

\begin{tw}\label{tw o jedynosci}
Assume that
\item$(B2)\qquad   \text{supp}\{\nu\}\subseteq [0,+\infty) \quad \text{and} \quad  \int_{1}^{\infty} y^{2} \nu (dy) <\infty.$\vskip2mm

\noindent If there exists a solution of the equation
\eqref{semigroup equation linear}  on the interval $[0,T^\ast]$
taking values in $H^{1,\gamma}_{+}$ then the solution is unique.
\end{tw}

\vskip4mm
\centerline{\underline{EQUIVALENT EQUATION}} \vskip2mm

\noindent {We pass now to the formulation of an equivalence result indicated in the introduction. It is of  independent interest and will
serve as the main technical tool in the majority of the proofs}.\vskip2mm

\noindent{A random field $r(t,x),\,\,t\in [0,T^\ast ], x\geq 0, \,\,$ is said to be a solution, in $L^{2,\gamma}$, respectively in $H^{1,\gamma}$,  to
the {\it integral equation}:
\begin{gather}\label{rownanie operatorwe na r}
r(t,x)=a(t,x)e^{\int_{0}^{t}J^{\prime}(\int_{0}^{t-s+x}\lambda(v)r(s,v)dv)\lambda(t-s+x)ds}, \quad x\geq 0,\,\,\,\,t\in[0,T^\ast ],
\end{gather}
where, for $x\geq 0,\,\, t\in (0,T^\ast],$
\begin{align}\label{wzor na tilde a}\nonumber
 a(t,x):=&r_0(t+x)e^{\int_{0}^{t}\lambda(t-s+x)dL(s)-\frac{q^2}{2}\int_{0}^{t}\lambda^2(t-s+x)ds}\\[2ex]
&\cdot\prod_{0\leq s\leq t}\left(1+\lambda(t-s+x)(L(s) -L(s-))\right)e^{-\lambda(t-s+x)(L(s) -L(s-))},
\end{align}
if $r(t,\cdot)\ ,\,t\in[0,T^\ast ],\,\,$ is $L^{2,\gamma}$, respectively  $H^{1,\gamma}$ valued, bounded and adapted process such that,
{for each $t\in[0,T^\ast ]$, the equation (\ref{rownanie operatorwe na r}) holds for almost all $x>0$, in the case of $L^{2,\gamma}$, and
for all ${x\geq 0}$, in the case of $H^{1,\gamma}$.}\vskip2mm

\noindent {The random field $a$ will be called  {\it the random factor} of the equation (\ref{rownanie operatorwe na r})}.
\begin{tw}\label{tw semigroup form  implies operator form} Let $r$ be a solution of \eqref{semigroup equation linear} in the state space
$H^{1,\gamma}_{+}$. Then $r(\cdot,\cdot)$ is a solution of \eqref{rownanie operatorwe na r} in $H^{1,\gamma}_{+}$.
\end{tw}
\vskip2mm

\noindent Under additional assumptions the converse result is true.}}

\begin{tw}\label{prop o rownowaznoesci rozwiazan}
Assume that  conditions $(\Lambda 0), (\Lambda 1)$  and
\begin{enumerate}[]
\item $(B0) \qquad\qquad   \int_{1}^{+\infty} y\nu(dy)<+\infty ,$
\end{enumerate}
are satisfied.
\begin{enumerate}[a)]
\item If
\begin{enumerate}[]
\item$(\Lambda 2)\qquad \qquad \lambda ,  \lambda^{\prime}  {\text{are bounded and  continuous on}}\,\,\,  \mathbb{R}_{+},$
\end{enumerate}
\noindent and $r(\cdot)$ is a bounded solution in  $L^{2,\gamma}_{+}$ of \eqref{rownanie operatorwe na r},
then $r(\cdot)$ is a c\`adl\`ag process in $L^{2,\gamma}_{+}$ and solves \eqref{semigroup equation linear}.
\item If
\begin{enumerate}[]
\item$(\Lambda 3)\qquad\qquad  \lambda ,\lambda^{\prime} ,\lambda^{\prime\prime},\,\, \text{are bounded and continuous on} \ \mathbb{R}_{+},$
\end{enumerate}
\begin{enumerate}[]
\item$(B2)\qquad\qquad   \text{supp}\{\nu\}\subseteq [0,+\infty) \quad \text{and} \quad  \int_{1}^{\infty} y^{2} \nu (dy) <\infty, $
\end{enumerate}
and  $r(\cdot)$ is a bounded solution in $H^{1,\gamma}_{+}$ of
\eqref{rownanie operatorwe na r}, then $r(\cdot)$ is  c\`adl\`ag in
$H^{1,\gamma}_{+}$ and solves \eqref{semigroup equation linear}.
\end{enumerate}
\end{tw}

As a consequence, equations \eqref{semigroup equation linear} and \eqref{rownanie operatorwe na r} are equivalent in $H^{1,\gamma}_{+}$, while each solution of \eqref{rownanie operatorwe na r} in $L^{2,\gamma}_{+}$ solves also \eqref{semigroup equation linear}.

\section{Proofs of the equivalence results}

 The proofs require representation of the solution in a natural and in a moving frame which is discussed in Section \ref{equivalence}. The proof of Theorem \ref{prop o rownowaznoesci rozwiazan} is technically rather involved. In particular it requires   auxiliary results concerned with the regularity of the random factor $a$  of the equation (\ref{rownanie operatorwe na r}).

\subsection{Equations in natural and moving frames}\label{equivalence}

{ We will need a result on  a relation
between the solution of the  equation \eqref{semigroup equation
linear} and its version in the natural frame.} To this end let us
consider two random fields $\{r(t,x), t,x\geq0\}$, $\{f(t,T), 0\leq
t\leq T<+\infty\}$ such that for each $x$ and each $T$ they admit
the following representation
\begin{align}\label{calkowa reprezentacja pola r}
r(t,x)&=r_0(t+x)+\int_{0}^{t}J^{\prime}
\Big(\int_{0}^{t-s+x}\delta(s,v)dv\Big)\delta(s,t-s+x)ds+\int_{0}^{t}\delta(s,t-s+x)dL(s),\\[2ex]\label{rownanie
na f przy HJM} f(t,T)&=f_0(T)+\int_{0}^{t}J^{\prime}
\left(\int_{s}^{T}\sigma(s,v)dv\right)\sigma(s,T)ds+\int_{0}^{t}\sigma(s,T)dL(s),
\end{align}
for some regular fields $\delta(\cdot,\cdot), \sigma(\cdot,\cdot)$
and initial conditions $r_0(\cdot), f_0(\cdot)$. We have the
following auxiliary lemma showing the relation between the dynamics
of $r$ and $f$ in the case when $ f(t,T)=r(t,T-t)$.
\begin{lem}\label{lemat o f i r}
\begin{enumerate}[a)]
\item Let $r$ be a random field given by \eqref{calkowa reprezentacja
pola r}. If $f(t,T):=r(t,T-t), 0\leq t\leq T<+\infty$ then $f$
satisfies \eqref{rownanie na f przy HJM} with
$\sigma(t,T):=\delta(t,T-t)$.
\item Let $f$ be a random field given by \eqref{rownanie na f przy
HJM}. If $r(t,x):=f(t,t+x), t,x\geq0$ then $r$ satisfies
\eqref{calkowa reprezentacja pola r} with
$\delta(t,x):=\sigma(t,t+x)$.
\end{enumerate}
\end{lem}
\noindent  {\bf Proof:} $(a)$ In virtue of
\eqref{calkowa reprezentacja pola r} we have {\small
\begin{align*}
f(t,T)&=r(t,T-t)=r_0(T)+\int_{0}^{t}J^{\prime}\left(\int_{0}^{T-s}\delta(s,v)dv\right)\delta(s,T-s)ds+
\int_{0}^{t}\delta(s,T-s)dL(s)\\[1ex]
&=r_0(T)+\int_{0}^{t}J^{\prime}\left(\int_{s}^{T}\delta(s,v-s)dv\right)\delta(s,T-s)ds+
\int_{0}^{t}\delta(s,T-s)dL(s)\\[1ex]
&=f_0(T)+\int_{0}^{t}J^{\prime}\left(\int_{s}^{T}\sigma(s,v)dv\right)\sigma(s,T)ds+\int_{0}^{t}\sigma(s,T)dL(s).
\end{align*}}
To get $(b)$ we can repeat the calculations in the reversed order.
\hfill$\square$

\subsection{Proof of Theorem \ref{tw semigroup form  implies
operator form}}

Let us consider the solution of \eqref{semigroup
equation linear} in a natural frame $f(t,T):=r(t,T-t)$, $0\leq t\leq
T<+\infty$. As convergence in $H^{1,\gamma}$ implies uniform
convergence on $[0,+\infty)$, see Lemma \ref{H-Lemma}, it follows
from the c\`adl\`ag property of $r$ in $H^{1,\gamma}_{+}$ that for
each $T>0$ the process $f(\cdot,T)$ is c\`adl\`ag on $[0,T]$. As $r$
satisfies \eqref{calkowa reprezentacja pola r} with
$\delta(t,x):=\lambda(t)r(t-,x)$, it follows from Lemma \ref{lemat o
f i r} that $f$ satisfies
\begin{align}\label{integral representation for f}\nonumber
f(t,T)=f_0(T)&+\int_{0}^{t}J^{\prime}\left(\int_{s}^{T}\lambda(v-s)f(s,v)dv\right)\lambda(T-s)f(s,T)ds\\[1ex]
&+\int_{0}^{t}\lambda(T-s)f(s-,T)dL(s).
\end{align}
Thus $f(\cdot,T)$ solves the Dol\'eans-Dade equation
\begin{gather*}
 df(t,T)=f(t-,T)\left[J^{\prime}\left(\int_{t}^{T}\lambda(v-t)f(t,v)dv\right)\lambda(T-t)dt+\lambda(T-t)dL(t)\right],
\end{gather*}
and admits the following representation, {see \cite{Protter},}
\begin{gather}\label{rrrownanie na f}
f(t,T)=\hat{a}(t,T)e^{\int_{0}^{t}J^{\prime}(\int_{s}^{T}\lambda(v-s)f(s,v)dv)\lambda(T-s)ds},
\end{gather}
with
\begin{gather*}
 \hat{a}(t,T):=f_0(T)e^{\int_{0}^{t}\lambda(T-s)dL(s)
 -\frac{1}{2}q^2\int_{0}^{t}\lambda^2(T-s)ds}\prod_{0\leq s\leq t}\Big(1+\lambda(T-s)\triangle L(s)\Big)e^{-\lambda(T-s)\triangle L(s)},
\end{gather*}
where
$$
\triangle L(s) = L(s)-L(s-),\,\,\,s\geq 0.
$$
Putting $T=t+x$, $x\geq0$ into \eqref{rrrownanie na f} and checking that $\hat{a}(t,t+x)=a(t,x)$ one obtains that $r$ satisfies \eqref{rownanie
operatorwe na r}. \hfill $\square$

\subsection{Proof of Theorem \ref{prop o rownowaznoesci rozwiazan}}
The proof is divided into two main steps establishing the regularity of the random factor $a$ and then the regularity of the nonlinear part
of(\ref{rownanie operatorwe na r}).
\subsubsection{Step 1. Regularity of the random factor of (\ref{rownanie operatorwe na r})}\label{Operator form for a global solution}
 Here we are dealing with the regularity of the random fields
\begin{align}\label{I_1}
I_1(t,x)&:=\int_{0}^{t}\lambda(t-s+x)dL(s), \quad t\in[0,T^\ast], \
x\geq0,
\\[1ex]\label{I_2}
I_2(t,x)&:=\prod_{0\leq s\leq t}\left(1+\lambda(t-s+x)\triangle
L(s)\right)e^{-\lambda(t-s+x)\triangle L(s)}, \qquad t\in[0,T^\ast],
x\geq0,
\end{align}
appearing in \eqref{wzor na tilde a}

\begin{prop}\label{prop o calkowaniu przez czesci}
Let $I_1$ be given by \eqref{I_1}. Assume that $(\Lambda 0), (\Lambda 1)$ are
satisfied.
\begin{enumerate}[i)]
\item If $(\Lambda 2)$ is satisfied then there exists a
version of the random field $I_1(t,x)$ which is bounded on
$[0,T^\ast]\times[0,+\infty)$ and for each $x\geq0$, the stochastic
integral $I_1(\cdot,x)$ is a c\`adl\`ag process.
\item If $(\Lambda 3)$ is satisfied then the above assertion is true
for the random field $\frac{\partial}{\partial x}I_1(t,x)$,
$t\in[0,T^\ast], \ x\geq0$.
\end{enumerate}
\end{prop}
{\bf Proof:} We will show $(i)$. The proof of $(ii)$ is similar. By
Proposition 9.16 of \cite{PeszatZabczyk} the integration by parts
formula holds
\begin{gather}\label{integration by parts}
I_1(t,x)=\int_{0}^{t}\lambda(t-s+x)dL(s)=\lambda(x)L(t)+\int_{0}^{t}\lambda^{\prime}(t-s+x)L(s)ds,
\quad t,x\geq 0.
\end{gather}
The integral on the right hand side of \eqref{integration by parts}
is continuous in $t$ as the convolution of two locally bounded
functions. Boundedness follows from the assumption $(\Lambda 2)$.
\hfill $\square$

\begin{prop}\label{prop o I_2}
Let $I_2$ be given by \eqref{I_2} and $(\Lambda 0), (\Lambda 1)$ be satisfied.
\begin{enumerate}[i)]
\item Then $I_2$ is a bounded field on
$[0,T^\ast]\times[0,+\infty)$ and for each $x\geq0$ the process
$I_2(\cdot,x)$ has c\`adl\`ag version.
\item If $(\Lambda 2)$ holds then the above assertion is true for
the field $\frac{\partial}{\partial x}I_2(t,x)$, $t\in[0,T^\ast], \
x\geq0$.
\end{enumerate}

\end{prop}
{\bf Proof:} Under $(\Lambda 0), (\Lambda 1)$ we can write $I_2$ in the form
\begin{gather*}
I_2(t,x)=\int_{0}^{t}\int_{-\frac{1}{\bar{\lambda}}}^{+\infty}
\big[\ln(1+\lambda(t-s+x)y)-\lambda(t-s+x)y\big]\pi(ds,dy),\quad
t\in[0,T^\ast], x\geq0,
\end{gather*}
where $\pi(ds,dx)$ stands for the jump measure of the process $L$.
Let us fix two numbers $a\leq0$ and $b>0$ such that
\begin{gather}\label{nierownosc dla skokow}
\mid \lambda(z)y\mid\leq\frac{1}{2}, \qquad z\geq0, \ y\in[a,b].
\end{gather}
Outside of the set $[0,T^\ast]\times[a,b]$ the measure $\pi$
consists of finite numbers of atoms only, so the fields
\begin{align*}
&\int_{0}^{t}\int_{-\frac{1}{\bar{\lambda}}}^{a}
\big[\ln(1+\lambda(t-s+x)y)-\lambda(t-s+x)y\big]\pi(ds,dy),\\[1ex]
&\int_{0}^{t}\int_{b}^{+\infty}
\big[\ln(1+\lambda(t-s+x)y)-\lambda(t-s+x)y\big]\pi(ds,dy), \quad
t\in[0,T^\ast], x\geq0,
\end{align*}
are obviously bounded and c\`adl\`ag in $t$. Thus required
properties of $I_2(t,x)$ are equivalent to those of the field
\begin{gather*}
J(t,x):=\int_{0}^{t}\int_{a}^{b}
\big[\ln(1+\lambda(t-s+x)y)-\lambda(t-s+x)y\big]\pi(ds,dy), \quad
t\in[0,T^\ast], x\geq0.
\end{gather*}
First we show boundedness. By \eqref{nierownosc dla skokow} we have
\begin{gather*}
\mid \ln(1+\lambda(z)y)-\lambda(z)y\mid\leq \lambda^2(z)y^2, \qquad
z\geq0, \ y\in[a,b],
\end{gather*}
and consequently
\begin{gather*}
\mid
J(t,x)\mid\leq\int_{0}^{t}\int_{a}^{b}\lambda^2(t-s+x)y^2\pi(ds,dy),
\qquad t\in[0,T^\ast], x\geq0.
\end{gather*}
Due to $(\Lambda 0), (\Lambda 1)$ boundedness of $J$ follows. Since
\begin{align*}
J^\prime_x(t,x)&=\int_{0}^{t}\int_{a}^{b}\lambda^\prime(t-s+x)y\Big[\frac{1}{1+\lambda(t-s+x)y}-1\Big]\pi(ds,dx)\\[1ex]
&=-\int_{0}^{t}\int_{a}^{b}\frac{\lambda^\prime(t-s+x)\lambda(t-s+x)}{1+\lambda(t-s+x)y}y^2\pi(ds,dx),
\end{align*}
in view of \eqref{nierownosc dla skokow}, the following estimation
holds
\begin{gather*}
\mid J^\prime_x(t,x)\mid=\int_{0}^{t}\int_{a}^{b}\frac{\mid
\lambda^\prime(t-s+x)\lambda(t-s+x)\mid}{\frac{1}{2}}y^2\pi(ds,dx).
\end{gather*}
Therefore, by $(\Lambda 2)$, boundedness of $J^\prime_x(t,x)$ and
thus also $\frac{\partial}{\partial x}I_2(t,x)$ follows.

Below we show c\`adl\`ag property for $I_2(\cdot,x)$. The proof for
$\frac{\partial}{\partial x}I_2(t,x)$ is the same. We will use the
following lemma.

\begin{lem}\label{lem o ciaglosci}
Assume that $\varphi(t,x,s,y)$, $(t,x)\in[0,T^\ast]\times
[0,+\infty)$, $(s,y)\in[0,T^\ast]\times [a,b]$, $a<b$, is a
continuous and bounded function such that
\begin{gather*}
\varphi(t,x,s,y)=0 \quad  \text{for} \ s\geq t, \ x\geq0, \ y\in[a,b],
\end{gather*}
and $\gamma$ is a finite measure on $[0,T^\ast]\times [a,b]$. Then
the function
\begin{gather*}
\Phi(t,x):=\int_{0}^{t}\int_{a}^{b}\varphi(t,x,s,y)\gamma(ds,dy),
\quad t\in[0,T^\ast], \ x\geq 0,
\end{gather*}
is continuous.
\end{lem}
{\bf Proof of Lemma \ref{lem o ciaglosci}:} By the assumptions,
\begin{gather*}
\Phi(t,x):=\int_{0}^{T^\ast}\int_{a}^{b}\varphi(t,x,s,y)\gamma(ds,dy),
\quad t\in[0,T^\ast], \ x\geq 0.
\end{gather*}
If $(t_n,x_n)\rightarrow(t,x)$ then
$\varphi(t_n,x_n,s,y)\rightarrow\varphi(t,x,s,y)$. Since $\varphi$
is bounded on $[0,T^\ast]\times
[0,+\infty)\times[0,T^\ast]\times[a,b]$ and $\gamma$ finite, the
result follows from the Lebesgue dominated convergence theorem.
\hfill $\square$

Now define a bounded and continuous function
\begin{gather*}
j(t,x,s,y):=\frac{1}{y^2}\big[\ln(1+\lambda(t-s+x)y)-\lambda(t-s+x)y\big],
\end{gather*}
for $(t,x)\in[0,T^\ast]\times [0,+\infty), (s,y)\in[0,T^\ast]\times
[a,b]$. Then
\begin{gather*}
J(t,x)=\int_{0}^{t}\int_{a}^{b}j(t,x,s,y)y^2\pi(ds,dy).
\end{gather*}
To use Lemma \ref{lem o ciaglosci} let us define

\[ \varphi(t,x,s,y):=
\begin{cases}
\quad j(t,x,s,y)-j(t,x,t,y), &s<t, x\geq0, y\in[a,b],\\
\quad 0,  &s\geq t, x\geq0, y\in[a,b],
   \end{cases}
\]
and $\gamma(ds,dy):=y^2\pi(ds,dy)$. Then
\begin{align*}
J(t,x)&=\int_{0}^{t}\int_{a}^{b}\varphi(t,x,s,y)y^2\pi(ds,dy)\\[1ex]
&+\int_{0}^{t}\int_{a}^{b}j(t,x,t,y)y^2\pi(ds,dy)\\[1ex]
&=\Phi(t,x)+\int_{a}^{b}j(t,x,t,y)y^2\pi([0,t],dy).
\end{align*}
The function $\Phi$ is continuous by Lemma \ref{lem o ciaglosci}
and thus $J(\cdot,x)$ is c\`adl\`ag for any $x\geq 0$. \hfill
$\square$ \vskip2mm

We will need one more result concerned with regularity of random
fields.
\begin{prop}\label{prop o funkcji cadlag}
 Let $h=h(x)\in L^{2,\gamma}$ and $H=H(t,x), t\in[0,T^\ast], x\geq 0$ be a function such that
 \begin{gather*}
  \sup_{(t,x)\in[0,T^\ast]\times[0,+\infty)}\mid H(t,x)\mid<+\infty,
 \end{gather*}
and $H(\cdot,x)$ is c\`adl\`ag for each $x\geq0$. Then the function
$\tilde{h}:[0,T^\ast]\longrightarrow L^{2,\gamma}$ defined by
\begin{gather*}
 \tilde{h}:=h(t+x)H(t,x)
\end{gather*}
is c\`adl\`ag in $L^{2,\gamma}$.
\end{prop}
{\bf Proof:} We have the following estimation
\begin{align*}
 \|\tilde{h}(t)&-\tilde{h}(s)\|^{2}_{L^{2,\gamma}}=\int_{0}^{+\infty}\mid h(t+x)H(t,x)-h(s+x)H(s,x)\mid^2e^{\gamma x}dx\\[2ex]
 &=\int_{0}^{+\infty}\mid h(s+x)[H(t,x)-H(s,x)]+[h(t+x)-h(s+x)]H(t,x)\mid^2e^{\gamma x}dx\\[2ex]
 &\leq 2 e^{-\gamma s}\int_{0}^{+\infty}e^{\gamma(x+s)}\mid h(s+x)\mid^2\mid H(t,x)-H(s,x)\mid^2dx+2C\|S_t(h)-S_s(h)\|^{2}_{L^{2,\gamma}},
\end{align*}
where $C=\sup_{(t,x)\in[0,T^\ast]\times[0,+\infty)}\mid H(t,x)\mid$.
Using the dominated convergence theorem we see that the limit for
the first integral when $s\rightarrow t$ exists and is equal to zero
when $s\downarrow t$. The second integral disappears when
$s\rightarrow t$ because the semigroup is strongly
 continuous in $L^{2,\gamma}$. Thus $\tilde{h}$ is a c\`adl\`ag function in $L^{2,\gamma}$.\hfill$\square$

\subsubsection{Step 2. A priori regularity of the solution}

\noindent Let us write \eqref{rownanie operatorwe na r} in the form
\begin{gather*}
 r(t,x)=r_0(t+x)B(t,x),
\end{gather*}
where $B(t,x):=b_1(t,x)I_2(t,x)b_2(t,x)$ and
\begin{align*}
 b_1(t,x)&:=e^{I_1(t,x)-\frac{q^2}{2}\int_{0}^{t}\lambda^2(t-s+x)ds},\\[2ex]
 b_2(t,x)&:=e^{\int_{0}^{t}J^{\prime}(\int_{0}^{t-s+x}\lambda(v)r(s,v)dv)\lambda(t-s+x)ds},
\end{align*}
where $I_1(t,x), I_2(t,x)$ are defined in \eqref{I_1} and
\eqref{I_2}.\\
\noindent
 $(a)$ First we will show that $r$ is c\`adl\`ag in $L^{2,\gamma}$.
We will show that $\sup_{(t,x)\in[0,T^\ast]\times[0,+\infty)}\mid
B(t,x)\mid<+\infty$ and $B(\cdot,x)$ is c\`adl\`ag for each $x$.
Then the assertion follows from Proposition \ref{prop o funkcji
cadlag}.

It follows from Proposition \ref{prop o calkowaniu przez czesci} and
Proposition \ref{prop o I_2} that  $(\Lambda 0), (\Lambda 1)$ and
$(\Lambda 2)$ imply that $b_1(t,x)$ and $I_2(t,x)$ are bounded and
c\`adl\`ag in $t$. It is clear that $b_2(\cdot,x)$ is continuous. By
$(\Lambda 1)$ and $(B0)$ the function $J^{\prime}$ is well defined
on $[0,+\infty)$. In view of \eqref{calkowalnosc rozwiazania} we
have
\begin{gather*}
0 \leq\int_{0}^{t-s+x}\lambda(v)r(s,v)dv\leq
\frac{\bar{\lambda}}{\sqrt{\gamma}}\sup_{t\in[0,T^\ast]}\|r(t)\|_{L^{2,\gamma}}.
\end{gather*}
and thus the inequality
\begin{gather*}
 \int_{0}^{t}J^{\prime}\left(\int_{0}^{t-s+x}\lambda(v)r(s,v)dv\right)\lambda(t-s+x)ds\leq \bar{\lambda}T^{\ast}J^{\prime}\left(\frac{\bar{\lambda}}{\sqrt{\gamma}}\sup_t\|r(t)\|_{L^{2,\gamma}}\right)\wedge 0,
\end{gather*}
holds. Thus $b_2(\cdot,\cdot)$ is bounded on
$[0,T^\ast]\times[0,+\infty)$.

Now we will argue that $r$ is a solution of \eqref{semigroup
equation linear}. Putting $x=T-t$ we see that the solution in the
natural frame satisfies
\begin{gather}
 f(t,T) =
\hat{a}(t,T)e^{\int_{0}^{t}J^{\prime}\left(\int_{s}^{T}\lambda(v-s)f(s,v)dv\right)\lambda(T-s)ds},\quad
0\leq t\leq T\leq T^\ast,
\end{gather}
where
\begin{gather*}
\hat{a}(t,T):=f_{0}(T)
 e^{\int_{0}^{t} \lambda(T-s) dL(s)-\frac{q^2}{2}\int_{0}^{t}\lambda^2(T-s)ds}\cdot\prod_{0\leq s\leq t}\Big(1+\lambda(T-s)\triangle L(s)\Big)e^{-\lambda(T-s)\triangle
 L(s)}.
\end{gather*}
For each fixed $T$ the process $f(\cdot,T)$ is a stochastic
exponential and thus admits the representation {\small
\begin{gather*}
 f(t,T)=f(0,T)+\int_{0}^{t}J^{\prime}\left(\int_{s}^{T}\lambda(v-s)f(s,v)dv\right)\lambda(T-s)f(s,T)ds
 +\int_{0}^{t}\lambda(T-s)f(s-,T)dL(s).
\end{gather*}}
Now the assertion follows from Lemma  \ref{lemat o f i r}.

\noindent $(b)$ To show that $r$ is  c\`adl\`ag in $H^{1,\gamma}$ we use the equality
\begin{gather*}
  \|r(t)-r(s)\|^{2}_{H^{1,\gamma}}=\|r(t)-r(s)\|^{2}_{L^{2,\gamma}}+\|r^{\prime}(t)-r^{\prime}(s)\|^{2}_{L^{2,\gamma}}.
\end{gather*}
Thus in view of $(a)$ it is enough to show that $r^{\prime}(t)$ is c\`adl\`ag in $L^{2,\gamma}$. Differentiating \eqref{rownanie operatorwe na r} yields
\begin{gather*}
r^{\prime}(t,x)=r^{\prime}_0(t+x)b(t,x)b_2(t,x)+r_0(t+x)b^{\prime}(t,x)b_2(t,x)+r_0(t+x)b(t,x)b^{\prime}_2(t,x)
\end{gather*}
where $b(t,x)=b_1(t,x)I_2(t,x)$. It follows from $(a)$ that
$r^{\prime}_0(t+x)b(t,x)b_2(t,x)$ is  c\`adl\`ag in $L^{2,\gamma}$.
In view of Proposition \ref{prop o calkowaniu przez czesci} and
Proposition \ref{prop o I_2}, $(\Lambda 2)$ and $(\Lambda 3)$ imply
that $b^{\prime}(t,x)$ is bounded and c\`adl\`ag in $t$, so
$r_0(t+x)b^{\prime}(t,x)b_2(t,x)$ is  c\`adl\`ag in $L^{2,\gamma}$.
To finish the proof we need to show that $b_2^{\prime}$ is bounded
and c\`adl\`ag in $t$. We have
\begin{align*}
b_2^{\prime}(t,x)=b_2(t,x)&\Big\{\int_{0}^{t}J^{\prime\prime}\left(\int_{0}^{t-s+x}\lambda(v)r(s,v)dv\right)\lambda^2(t-s+x)r(s,t-s+x)ds\\[2ex]
&+\int_{0}^{t}J^{\prime}\left(\int_{0}^{t-s+x}\lambda(v)r(s,v)dv\right)\lambda^{\prime}(t-s+x)ds\Big\}.
\end{align*}
The assumptions $(\Lambda 1)$ and (B2) guarantee that
$J^{\prime\prime}$ is continuous on $[0,+\infty)$ and thus locally
bounded. In view of Lemma \ref{H-Lemma} we obtain
\begin{align*}
\sup_{(t,x)\in[0,T^\ast]\times[0,+\infty)}\int_{0}^{t}J^{\prime\prime}\left(\int_{0}^{t-s+x}\lambda(v)r(s,v)dv\right)\lambda^2(t-s+x)r(s,t-s+x)ds\\[2ex]
\leq\bar{\lambda}^2\sup_{z\in[0,\frac{\bar{\lambda}}{\sqrt{\gamma}}\sup_t\|r(t)\|_{L^{2,\gamma}_{+}}]}\mid
J^{\prime\prime}(z)\mid \cdot
2T^\ast\left(\frac{1}{\gamma}\right)^{\frac{1}{2}}\sup_{t}\|r(t)\|_{H^{1,\gamma}_{+}}.
\end{align*}
By monotonicity of $J^{\prime}$ and boundedness of
$\lambda^{\prime}$ one gets
\begin{align*}
\sup_{(t,x)\in[0,T^\ast]\times[0,+\infty)}\int_{0}^{t}J^{\prime}\left(\int_{0}^{t-s+x}\lambda(v)r(s,v)dv\right)\lambda^{\prime}(t-s+x)ds\\[2ex]
\leq T^\ast\sup_{x\geq0}\mid\lambda^{\prime}(x)\mid
J^{\prime}\left(\frac{\bar{\lambda}}{\sqrt{\gamma}}\sup_t\|r(t)\|_{L^{2,\gamma}}\right),
\end{align*}
and boundedness of $b^\prime_2$ follows. The proof that $r$ solves
\eqref{semigroup equation linear} is the same as in $(a)$. \hfill
$\square$

\section{Proofs of necessary conditions for existence in
$H^{1,\gamma}_{+}$}\label{Necessary conditions for existence}

\noindent{\bf Proof of Theorem \ref{tw o
eksplozjach}:} Assume to the contrary that $r$ is a global solution
of \eqref{semigroup equation linear} on $[0,T^\ast]$ in the space
$H^{1,\gamma}_{+}$. In view of Lemma \ref{lemat o f i r} the
solution in a moving frame $f(t,T)=r(t,T-t), 0\leq t\leq T\leq
T^\ast$ satisfies
\begin{align}\nonumber
f(t,T)=f_0(T)&+\int_{0}^{t}J^{\prime}\left(\int_{s}^{T}\lambda(v-s)f(s,v)dv\right)\lambda(T-s)f(s,T)ds\\[1ex]\label{row f}
&+\int_{0}^{t}\lambda(T-s)f(s-,T)dL(s),
\end{align}
{which is the equation  studied in \cite{BarZab}.}
Assumptions $(\Lambda0),(\Lambda1),(\Lambda3),(B0)$ imply the
conditions $(A1)-(A4)$ in \cite{BarZab}. 

\noindent We check $f$ is as regular as required in  \cite{BarZab}. Since $r$ is adapted and
c\`adl\`ag in $H^{1,\gamma}_{+}$, it follows that
\begin{enumerate}[(a)]
 \item $f(\cdot,T)$ is adapted and c\`adl\`ag for each $T\in[0,T^\ast]$,
 \item $f(t,\cdot)$ is continuous.
 \end{enumerate}
Using Lemma \ref{H-Lemma} and the fact that $r$ is bounded on $[0,T^\ast]$, as a c\`adl\`ag process in $H^{1,\gamma}_{+}$, we obtain
\begin{gather*}
 \sup_{t\in[0,T^\ast], x\geq0}\mid r(t,x)\mid=\sup_{t\in[0,T^\ast]}\sup_{x\geq 0}\mid r(t,x)\mid\leq 2\left(\frac{1}{\gamma}\right)^{\frac{1}{2}}\sup_{t\in[0,T^\ast]}\|r\|_{H^{1,\g}_{+}}<+\infty,
\end{gather*}
which clearly implies that
\begin{gather*}
 (c) \qquad \sup_{0\leq t\leq T\leq T^\ast} f(t,T)<+\infty.
\end{gather*}
It follows, however,  from Theorem 3.4 in \cite{BarZab} that, under (B3),  
for sufficiently large $k>0$ there is no solution of \eqref{row f}
in the class of random fields satisfying $(a)-(c)$. Hence   a
contradiction.

\hfill$\square$

\section{Proofs of existence of global and strong
solutions}\label{Existence of global solutions}

In view of Theorem \ref{prop o rownowaznoesci rozwiazan} we can
examine equation \eqref{rownanie operatorwe na r} instead directly
\eqref{semigroup equation linear}. Let us begin with clarifying of
the general idea of examining the problem of existence of solution
to the equation \eqref{rownanie operatorwe na r}. Define the
operator $\mathcal{K}$, acting on functions of two variables, by
\begin{gather}
\mathcal{K}(h)(t,x)=
a(t,x)e^{\int_{0}^{t}J^{\prime}\left(\int_{0}^{t-s+x}\lambda(v)h(s,v)dv\right)\lambda(s,t-s+x)ds},
\quad x\geq 0,\,\,\,\,t\in [0,T^\ast ],
\end{gather}
where $a(t,x)$ is given by \eqref{wzor na tilde a}. Then the
equation \eqref{rownanie operatorwe na r} can be compactly written
in the form
$$
r(t,x)=\mathcal{K}(r)(t,x),\,\,\,\quad t\in [0,T^{*}] ,\quad  x\geq0
\ .
$$
The problem of existence of solutions will be examined via
properties of the iterative sequence of random fields
\begin{gather}\label{schemat iteracyjny}
h_0\equiv0, \qquad h_{n+1}:=\mathcal{K}h_n, \qquad n=1,2,\dots \ .
\end{gather}
Let us write $a$ in the form $a(t,x)=r_0(t+x)b(t,x)$. It follows
from Proposition \ref{prop o calkowaniu przez czesci} and
Proposition \ref{prop o I_2} that under $(\Lambda 2)$ the field $b$
is bounded, i.e.
\begin{gather}\label{ograniczonosc tilde b}
\sup_{t\in[0,T^\ast], x\geq0} b(t,x)<\bar{b},
\end{gather}
where $\bar{b}=\bar{b}(\omega)>0$. It can be shown by induction that
if $r_0\in L^{2,\gamma}_{+}$ then $h_n(t)$ is a bounded process in
$L^{2,\gamma}_{+}$ for each $n$. Indeed assume that for $h_n$ and
show for $h_{n+_1}$. In view of \eqref{ograniczonosc tilde b} and
the estimate \eqref{calkowalnosc rozwiazania} in Appendix, we have
\begin{align*}
h_{n+1}(t,x)&\leq
r_0(t+x)\ \bar{b}\ e^{\bar{\lambda}\int_{0}^{t}\mid J^{\prime}(\int_{0}^{t-s+x}\lambda(v)h_n(s,v)dv)\mid ds}\\[2ex]
&\leq r_0(t+x) \ \bar{b} \ e^{\bar{\lambda}T^\ast\big|
J^{\prime}(\frac{\bar{\lambda}}{\sqrt{\gamma}}\sup{t}\parallel
h_n(t)\parallel_{L^{2,\gamma}})\big|},
\end{align*}
and thus $h_{n+1}(t)$ is bounded in $L^{2,\gamma}_{+}$. It follows
from the assumption $\underline{\lambda}>0$ and the fact that
$J^{\prime}$ is increasing that the sequence $\{h_n\}$ is
monotonically increasing and thus there exists
$\bar{h}:[0,T^\ast]\times[0,+\infty)\longrightarrow \mathbb{R}_{+}$
such that
\begin{gather}\label{granica w schemacie iteracyjnym}
\lim_{n\rightarrow +\infty}h_n(t,x)=\bar{h}(t,x), \qquad 0\leq t\leq
T^\ast, x\geq0.
\end{gather}
Passing to the limit in \eqref{schemat iteracyjny}, by the monotone
convergence, we obtain
\begin{gather*}
\bar{h}(t,x)=\mathcal{K}h(t,x), \qquad 0\leq t\leq T^\ast, x\geq0.
\end{gather*}
It turns out that properties of the field $\bar{h}$ strictly depend
on the growth of the function $J^{\prime}$. In the sequel we show
that if (B4) holds then $\bar{h}(t)$ is a bounded process in
$L^{2,\gamma}_{+}$, i.e. $\bar{h}(t), t\in[0,T^\ast]$ is a
non-exploding solution of \eqref{rownanie operatorwe na r} in
$L^{2,\gamma}_{+}$. Additional assumptions guarantee that
$\bar{h}(t)$ is bounded in $H^{1,\gamma}_{+}$ and that the solution
is unique.

\noindent Before presenting the proof we establish  an auxiliary
result.
\begin{prop}\label{prop pomocniczy}
Assume that $J^{\prime}$ satisfies (B4). If $r_0\in L^{2,\gamma}_{+}$ then there exists a positive
constant $c_1$ such that if
\begin{gather*}
\sup_{t\in[0,T^\ast]}\|h(t)\|_{L^{2,\gamma}_{+}}\leq c_1
\end{gather*}
then
\begin{gather*}
\sup_{t\in[0,T^\ast]}\|\mathcal{K}h(t)\|_{L^{2,\gamma}_{+}}\leq c_1.
\end{gather*}
\end{prop}
{\bf Proof}: By \eqref{calkowalnosc rozwiazania} in Appendix and
\eqref{ograniczonosc tilde b}, for any $t\in[0,T^\ast]$, we have
\begin{align*}
\|\mathcal{K}h(t,\cdot)\|_{L^{2,\gamma}_{+}}^2&=\int_{0}^{+\infty}|r_0(t+x)b(t,x)|^2e^{2\int_{0}^{t}J^{\prime}(\int_{0}^{t-s+x}\lambda(v)h(s,v)dv)\lambda(t-s+x)ds}
e^{\gamma x}dx\\[2ex]
&\leq \bar{b}^2\int_{0}^{+\infty}|r_0(t+x)|^2e^{2
J^{\prime}\left(\frac{\bar{\lambda}}{\sqrt{\gamma}}\cdot\sup_{t}\|h(t)\|_{L^{2,\gamma}_{+}}\right)\int_{0}^{t}\lambda(t-s+x)ds}e^{\gamma
x}dx\\[2ex]
&\leq \bar{b}^2\cdot
\|r_0\|^2_{L^{2,\gamma}_{+}}\cdot\sup_{s\in[0,t], x\geq0}
e^{2J^{\prime}\left(\frac{\bar{\lambda}}{\sqrt{\gamma}}\cdot\sup_t\|h(t)\|_{L^{2,\gamma}_{+}}\right)\int_{0}^{t}\lambda(t-s+x)ds}.
\end{align*}
This implies
\begin{gather*}
\sup_{t}\|\mathcal{K}h(t)\|_{L^{2,\gamma}_{+}}\leq \bar{b} \cdot
\|r_0\|_{L^{2,\gamma}_{+}} \cdot \sup_{t\in[0,T^\ast], s\in[0,t],
x\geq0}e^{
J^{\prime}\left(\frac{\bar{\lambda}}{\sqrt{\gamma}}\cdot\sup_{t}\|h(t)\|_{L^{2,\gamma}_{+}}\right)\int_{0}^{t}\lambda(t-s+x)ds},
\end{gather*}
and thus it is enough to find constant $c_1$ such that
\begin{gather}\label{warunek na c}
\ln\left(\bar{b}\cdot\|r_0\|_{L^{2,\gamma}_{+}}\right)+
\sup_{t\in[0,T^\ast], s\in[0,t],
x\geq0}J^{\prime}\left(\frac{\bar{\lambda}c_1}{\sqrt{\gamma}}\right)\int_{0}^{t}\lambda(t-s+x)ds\leq
\ln c_1.
\end{gather}
If $J^{\prime}(z)\leq 0$ for each $z\geq0$ then we put
$c_1=\bar{b}\cdot\|r_0\|_{L^{2,\gamma}_{+}}$. If $J^{\prime}$ takes
positive values then it is enough to find large $c_1$ such that
\begin{gather*}
\ln\left(\bar{b}\cdot\|r_0\|_{L^{2,\gamma}_{+}}\right)\leq \ln c_1 -
\bar{\lambda}T^\ast
J^{\prime}\left(\frac{\bar{\lambda}c_1}{\sqrt{\gamma}}\right).
\end{gather*}
Existence of such $c_1$ is a consequence of (B4).\hfill$\square$\\

\noindent {\bf Proof of Theorem \ref{tw o istnieniu}}: Since
$\bar{h}(\cdot,x)$ is adapted for each $x\geq0$ as a pointwise
limit, we only need to show that $\bar{h}(t)$ is a bounded process
in $L^{2,\gamma}_{+}$, resp. $H^{1,\gamma}_{+}$. Then $\bar{h}$
solves \eqref{semigroup equation linear} in virtue of Theorem
\ref{prop o rownowaznoesci rozwiazan}.

\noindent $(a)$  Let $c_1$ be a constant given by Proposition
\ref{prop pomocniczy}. By the Fatou lemma we have
\begin{gather*}
\sup_{t\in[0,T^\ast]}\int_{0}^{+\infty}\mid \bar{h}(t,x)\mid^2
e^{\gamma x}dx\leq \sup_{t\in[0,T^\ast]}\underset{n\rightarrow
+\infty}{\lim \inf}\int_{0}^{+\infty}\mid h_n(t,x)\mid^2 e^{\gamma
x}dx\leq c_1^2,
\end{gather*}
and hence $\bar{h}(t)$ is bounded in $L^{2,\gamma}_{+}$.

\noindent $(b)$ In view of $(a)$ we need to show that
$h^{\prime}_x(t)$ is bounded in $L^{2,\gamma}$. Differentiating the
equation $\bar h=\mathcal{K}\bar{h}$ yields
\begin{gather*}
\bar{h}^{\prime}(t,x)=r_0^{\prime}(t+x)b(t,x)F_1(t,x)+r_0(t+x)b^{\prime}_x(t,x)F_1(t,x)+r_0(t+x)b(t,x)F_1(t,x)F_2(t,x),
\end{gather*}
where
\begin{align*}
F_1(t,x)&:=e^{\int_{0}^{t}J^{\prime}(\int_{0}^{t-s+x}\lambda(v)\bar{h}(s,v)dv)\lambda(t-s+x)ds},\\[2ex]
F_2(t,x)&:=\int_{0}^{t}J^{\prime\prime}\left(\int_{0}^{t-s+x}\lambda(v)\bar{h}(s,v)dv\right)\lambda^2(t-s+x)\bar{h}(s,t-s+x)ds\\[2ex]
&+\int_{0}^{t}J^{\prime}\left(\int_{0}^{t-s+x}\lambda(v)h(s,v)dv\right)\lambda^{\prime}_{x}(t-s+x)ds.
\end{align*}
Assumption $(\Lambda 3)$ implies that $b(\cdot,\cdot)$ and
$b^{\prime}_x(\cdot,\cdot)$ are bounded on
$(t,x)\in[0,T^\ast]\times[0,+\infty)$. Since $r_0\in
H^{1\gamma}_{+}$, it is enough to show that
\begin{gather*}
\sup_{t\in[0,T^\ast], x\geq 0}F_1(t,x)<+\infty, \quad
\sup_{t\in[0,T^\ast], x\geq 0}F_2(t,x)<+\infty.
\end{gather*}
\noindent We have
\begin{gather*}
\sup_{t\in[0,T^\ast], x\geq 0}F_1(t,x)\leq e^{\big|
J^{\prime}\left(\frac{\bar{\lambda}}{\sqrt{\gamma}}\sup_t\parallel
\bar{h}(t)\parallel_{L^{2,\gamma}_{+}}\right)\big|\bar{\lambda}T^\ast}<+\infty.
\end{gather*}

It follows from Proposition \ref{tw o subordynatorze} that (B4)
excludes Wiener part of the noise as well as negative jumps. Thus
$J^{\prime\prime}$ reduces to the form
$J^{\prime\prime}(z)=\int_{0}^{+\infty}y^2e^{-zy}\nu(dy)$ and $0\leq
J^{\prime\prime}(0)<+\infty$ due to the assumption (B2). Since
$J^{\prime\prime}$ is decreasing, the following estimation holds
\begin{align*}
\sup_{t\in[0,T^\ast], x\geq 0}F_2(t,x)\leq&
J^{\prime\prime}(0)T^\ast\bar{\lambda}^2\sup_{t\in[0,T^\ast], x\geq
0}\int_{0}^{t}\bar{h}(s,t-s+x)ds\\[2ex]
&+T^\ast\big|
J^{\prime}\left(\frac{\bar{\lambda}}{\sqrt{\gamma}}\sup_t\parallel
\bar{h}(t)\parallel_{L^{2,\gamma}_{+}}\right)\big|\cdot\sup_{x\geq0}\lambda^{\prime}(x),
\end{align*}
and it is enough to show that $\bar{h}$ is bounded on $\{(t,x),
t\in[0,T^\ast], x\geq0\}$. In view of the fact that
$\bar{h}=\mathcal{K}\bar{h}$  we obtain
\begin{gather*}
\sup_{t\in[0,T^\ast], x\geq
0}\bar{h}(t,x)\leq\sup_{x\geq0}r_0(x)\cdot\sup_{t\in[0,T^\ast],
x\geq 0}b(t,x)\cdot
e^{\big|J^{\prime}\left(\frac{1}{\sqrt{\gamma}}\sup_{t}\parallel\bar{h}(t)\parallel_{L^{2,\gamma}_{+}}\right)\big|\bar{\lambda}T^\ast}<+\infty.
\end{gather*}
\hfill $\square$

\noindent
{\bf Proof of Theorem \ref{tw o silnym rozwiazaniu}:}\\
\noindent Let $r$ be a solution given by Theorem \ref{tw o
istnieniu}$(b)$. Then, by Theorem \ref{tw semigroup form  implies
operator form}, $r$ solves \eqref{rownanie operatorwe na r}. We will
show that the assumption $\lambda(\cdot)=\lambda$ implies that $r$
is a solution of equation \eqref{general equation}. Differentiating
\eqref{rownanie operatorwe na r} yields {\small
\begin{align}\label{pochodna r}\nonumber
\frac{\partial}{\partial x} r(t,x)&= e^{\lambda
L_t-\frac{q^2\lambda^2}{2}}\prod(1+\lambda\triangle
L_s)e^{-\lambda\triangle
L_s}\cdot\\[2ex]\nonumber
&\cdot\bigg(r^{\prime}_0(t+x)e^{\lambda\int_{0}^{t}J^{\prime}(\lambda\int_{0}^{t-s+x}r(s,v)dv)ds}+r_0(t+x)e^{\lambda\int_{0}^{t}J^{\prime}(\lambda\int_{0}^{t-s+x}r(s,v)dv)ds}\cdot\\[2ex]\nonumber
&\phantom{aaaaaaaaaaaaaaaaaaaaaa}
\cdot\lambda^2\int_{0}^{t}J^{\prime\prime}\Big(\lambda\int_{0}^{t-s+x}r(s,v)dv\Big)\cdot
r(s,t-s+x)ds\bigg)\\[2ex]\nonumber
&=r(t,x)\frac{r^{\prime}_0(t+x)}{r_0(t+x)}+r(t,x)\lambda^2\int_{0}^{t}J^{\prime\prime}\Big(\lambda\int_{0}^{t-s+x}r(s,v)dv\Big)\cdot
r(s,t-s+x)ds\\[2ex]
&=r(t,x)\bigg[ \
\frac{r^{\prime}_0(t+x)}{r_0(t+x)}+\lambda^2\int_{0}^{t}J^{\prime\prime}\Big(\lambda\int_{0}^{t-s+x}r(s,v)dv\Big)\cdot
r(s,t-s+x)ds\bigg].
\end{align}}
For $Z_1$, $Z_2$ defined by
\begin{align*}
Z_1(t)&:=e^{\lambda
L_t-\frac{q^2\lambda^2}{2}}\prod(1+\lambda\triangle
L_s)e^{-\lambda\triangle
L_s},\\[2ex]
Z_2(t,x)&:=r_0(t+x)e^{\lambda\int_{0}^{t}J^{\prime}(\lambda\int_{0}^{t-s+x}r(s,v)dv)ds},
\end{align*}
we have SDEs of the form {\small
\begin{align*}
dZ_1(t)&=Z_1(t-)\lambda dL(t)\\[2ex]
dZ_2(t,x)&=\bigg\{r^{\prime}_0(t+x)e^{\lambda\int_{0}^{t}J^{\prime}(\lambda\int_{0}^{t-s+x}r(s,v)dv)ds}+r_0(t+x)e^{\lambda\int_{0}^{t}J^{\prime}(\lambda\int_{0}^{t-s+x}r(s,v)dv)ds}\cdot\\[2ex]
&\cdot\Big[\lambda J^{\prime}\Big(\lambda\int_{0}^{x}r(t,v)dv\Big)+\lambda^2\int_{0}^{t}J^{\prime\prime}\Big(\lambda\int_{0}^{t-s+x}r(s,v)dv\Big)r(s,t-s+x)ds\Big]\bigg\}dt\\[2ex]
&=\bigg\{\frac{r^{\prime}_0(t+x)}{r_0(t+x)}Z_2(t,x)+Z_2(t,x)\Big[\lambda J^{\prime}\Big(\lambda\int_{0}^{x}r(t,v)dv\Big)+\\[2ex]
&\phantom{aaaaaaaaaaaaaaaaaaaaaaaa}+\lambda^2\int_{0}^{t}J^{\prime\prime}\Big(\lambda\int_{0}^{t-s+x}r(s,v)dv\Big)r(s,t-s+x)ds\Big]\bigg\}dt\\[2ex]
&=\bigg\{Z_2(t,x)\bigg[\frac{r^{\prime}_0(t+x)}{r_0(t+x)}
+\lambda J^{\prime}\Big(\lambda\int_{0}^{x}r(t,v)dv\Big)+\\[2ex]
&\phantom{aaaaaaaaaaaaaaaaaaaaaaaa}+\lambda^2\int_{0}^{t}J^{\prime\prime}\Big(\lambda\int_{0}^{t-s+x}r(s,v)dv\Big)r(s,t-s+x)ds\bigg]\bigg\}dt.
\end{align*}}
Using the formulas above, we obtain SDE for $r(t,x)$: {\small
\begin{align*}
dr(t,x)&=d\Big(Z_1(t)Z_2(t,x)\Big)=Z_1(t)dZ_2(t,x)+Z_2(t,x)dZ_1(t)\\[2ex]
&=Z_1(t)Z_2(t,x)\bigg[\frac{r^{\prime}_0(t+x)}{r_0(t+x)}
+\lambda J^{\prime}\Big(\lambda\int_{0}^{x}r(t,v)dv\Big)+\\[2ex]
&\phantom{aaaaaaaaaaaaaaaaaaaaaa}+\lambda^2\int_{0}^{t}J^{\prime\prime}\Big(\lambda\int_{0}^{t-s+x}r(s,v)dv\Big)r(s,t-s+x)ds\bigg]dt\\[2ex]
&+Z_2(t,x)Z_1(t-)\lambda dL(t)\\[2ex]
&=r(t,x)\bigg[\frac{r^{\prime}_0(t+x)}{r_0(t+x)}
+\lambda^2\int_{0}^{t}J^{\prime\prime}\Big(\lambda\int_{0}^{t-s+x}r(s,v)dv\Big)r(s,t-s+x)ds\bigg]dt\\[2ex]
&+\lambda r(t,x)J^{\prime}\Big(\lambda \int_{0}^{x}r(t,v)dv\Big)dt+\lambda r(t-,x)dL(t)\\[2ex]
&\overset{by \eqref{pochodna r}}{=}\frac{\partial}{\partial x}
r(t,x)dt+\lambda
J^{\prime}\Big(\lambda\int_{0}^{x}r(t,v)dv\Big)r(t,x)dt+\lambda
r(t-,x)dL(t),
\end{align*}}
which is \eqref{general equation}. \hfill $\square$

\section{Proof of the uniqueness of the
solutions in $H^{1,\gamma}_{+}$}\label{Uniqueness}

Before presenting the proof of Theorem \ref{tw o jedynosci} we
establish an auxiliary result.

\begin{prop}\label{prop Gronwall}
Let $d:[0,T^\ast]\times[0,+\infty)\longrightarrow\mathbb{R}_{+}$ be a bounded
function satisfying
\begin{gather}\label{wzor w Gronwallu}
d(t,x)\leq C \int_{0}^{t}\int_{0}^{t-s+x}d(s,v)dvds,
\end{gather}
where $C>0$ is a fixed constant. Then $d(t,x)=0$ for all $(t,x)\in
[0,T^\ast]\times[0,+\infty)$.
\end{prop}
{\bf Proof:} Let $d$ be bounded by $M>0$ on
$[0,T^\ast]\times[0,+\infty)$. Let us define a new function
\begin{gather*}
\bar{d}(u,w):=d(u,w-u); \qquad u\in[0,T^\ast], w\geq u.
\end{gather*}
It is clear that $d\equiv0$ on $[0,T^\ast]\times[0,+\infty)$ if and
only if $\bar{d}\equiv 0$ on the set $\{(u,w): u\in[0,T^\ast], w\geq
u\}$. Let us notice that \eqref{wzor w Gronwallu} implies that
\begin{align*}
\bar{d}(u,w)&=d(u,w-u)\leq C\int_{0}^{u}\int_{0}^{w-s}d(s,y)dy
ds\\[2ex]
&= C \int_{0}^{u}\int_{s}^{w}d(s,z-s)dz ds=C
\int_{0}^{u}\int_{s}^{w}\bar{d}(s,z)dz ds.
\end{align*}
Using this inequality we will show by induction that
\begin{gather}\label{indukcja 2}
\bar{d}(u,w)\leq MC^n\frac{(uw)^n}{(n!)^2}, \qquad n=0,1,2,...\ .
\end{gather}
Then letting $n\rightarrow0$ we have $\bar{d}(t,x)=0$. The formula
\eqref{indukcja 2} is valid for $n=0$. Assume that it is true for
$n$ and show for $n+1$:
\begin{align*}
\bar{d}(u,w)&\leq C\int_{0}^{u}\int_{s}^{w}MC^{n}\frac{(sz)^n}{(n!)^2}dzds=MC^{n+1}\frac{1}{(n!)^2}\int_{0}^{u}s^n(\int_{s}^{w}z^n dz)ds\\[2ex]
&=
MC^{n+1}\frac{1}{(n!)^2}\int_{0}^{u}s^n\left(\frac{w^{n+1}-s^{n+1}}{n+1}\right)ds\leq
MC^{n+1}\frac{1}{(n!)^2}\int_{0}^{u}s^n\frac{w^{n+1}}{n+1}ds\\[2ex]
&=MC^{n+1}\frac{1}{(n!)^2}\frac{u^{n+1}}{(n+1)}\frac{w^{n+1}}{(n+1)}=MC ^{n+1}\frac{(uw)^{n+1}}{((n+1)!)^2}.
\end{align*}
\hfill $\square$

\noindent {\bf Proof of Theorem \ref{tw o jedynosci}:} Assume that
$r_1, r_2$ are two solutions of the equation \eqref{semigroup
equation linear} in $H^{1,\gamma}_{+}$. Then they are bounded
processes in $H^{1,\gamma}$ and, in view of Theorem \ref{tw
semigroup form  implies operator form}, satisfy \eqref{rownanie
operatorwe na r}. Define
\begin{gather*}
d(t,x):=\mid r_{1}(t,x)-r_2(t,x)\mid, \qquad 0\leq t\leq T^\ast,
x\geq0.
\end{gather*}
Denote  $B:=\sup_{t\in[0,T^\ast],x\geq0}b(t,x)$. The following
estimation holds
\begin{align*}
d(t,x)&\leq
r_0(t+x)b(t,x)\left[e^{\int_{0}^{t}J^{\prime}(\int_{0}^{t-s+x}\lambda(s,v)r_1(s,v)dv)\lambda(s,t-s+x)ds}
+e^{\int_{0}^{t}J^{\prime}(\int_{0}^{t-s+x}\lambda(s,v)r_2(s,v)dv)\lambda(s,t-s+x)ds}\right]\\[2ex]
&\leq \sup_{x\geq0}r_0(x)\cdot B\cdot\left[e^{\bar{\lambda}T^\ast
\big|J^{\prime}(\frac{\bar{\lambda}}{\sqrt{\gamma}}\sup_{t}\|r_1(t)\|_{L^{2,\gamma}_{+}})\big|}
+e^{\bar{\lambda}T^\ast
\big|J^{\prime}(\frac{\bar{\lambda}}{\sqrt{\gamma}}\sup_t\|r_2(t)\|_{L^{2,\gamma}_{+}})\big|}
\right]<+\infty,
\end{align*}
and thus $d$ is bounded on $[0,T^\ast]\times[0,+\infty)$. In view of
the inequality $\mid e^{x}-e^{y}\mid\leq e^{x\vee y}\mid x-y\mid; \
x,y\geq0$ and the fact that $J^{\prime\prime}$ is decreasing with
$0\leq J^{\prime\prime}(0)<+\infty$ due to assumption (B2), we have
{\small
\begin{align*}
&d(t,x)\leq
\sup_{x\geq0}r_0(x)\cdot Be^{\max\Big\{\int_{0}^{t}J^{\prime}\Big(\int_{0}^{t-s+x}\lambda(s,v)r_1(s,v)dv\Big)\lambda(s,t-s+x)ds;\int_{0}^{t}J^{\prime}\Big(\int_{0}^{t-s+x}\lambda(s,v)r_2(s,v)dv\Big)\lambda(s,t-s+x)ds\Big\}}\cdot\\[2ex]
&\cdot\left|\int_{0}^{t}J^{\prime}\left(\int_{0}^{t-s+x}\lambda(s,v)r_1(s,v)dv\right)\lambda(s,t-s+x)ds-\int_{0}^{t}J^{\prime}\left(\int_{0}^{t-s+x}\lambda(s,v)r_2(s,v)dv\right)\lambda(s,t-s+x)ds\right|\\[2ex]
&\leq \sup_{x\geq0}r_0(x)\cdot Be^{\bar{\lambda}T^\ast\max\Big\{
\big|J^{\prime}\big({
\frac{\bar{\lambda}}{\sqrt{\gamma}}\sup_t\|r_1(t)\|_{L^{2,\gamma}_{+}}}\big)\big|;
\big|J^{\prime}\big({
\frac{\bar{\lambda}}{\sqrt{\gamma}}\sup_t\|r_2(t)\|_{L^{2,\gamma}_{+}}}\big)\big|\Big\}}\cdot\\[2ex]
&\cdot
J^{\prime\prime}(0)\bar{\lambda}^2\int_{0}^{t}\int_{0}^{t-s+x}\mid
r_1(s,v)-r_2(s,v)\mid dv ds = C \int_{0}^{t}\int_{0}^{t-s+x}d(s,v)dv
ds, \qquad (t,x)\in [0,T^\ast]\times[0,+\infty).
\end{align*}}
It follows from Proposition \ref{prop Gronwall} that $r_1=r_2$ on
$[0,T^\ast]\times[0,+\infty)$. \hfill$\square$

\section{Appendix}

\subsection{HJM approach to the bond market}\label{HJM approach to the bond market}

Let $P(t,T)$ denote a price at time $t\geq0$ of a bond paying $1$ unit of money to its holder at time $T\geq t$. The prices $P(\cdot,T)$ are processes
defined on a fixed filtered probability space $(\Omega, \mathcal{F}_{t, t\geq0},P)$. The forward rate $f$ is a random field defined by the formula
\begin{gather*}
P(t,T)= e^{-\int_{t}^{T}f(t,u)du}, \qquad 0\leq t\leq T\leq T^\ast.
\end{gather*}
The prices of all bonds traded on the market are thus determined by the forward rate $f(t,T), 0\leq t\leq T<+\infty$ and thus the starting point in the
bond market description is specifying the dynamics of $f$. In this paper we consider the following stochastic differentials
\begin{gather}\label{rownanie na f}
df(t,T)=\alpha(t,T)dt+\sigma(t,T)dL(t),\qquad 0\leq t\leq T,
\end{gather}
where $L$ is a L\'evy process. The equation above can be viewed as a system of infinitely many equations parameterized by $0\leq T<+\infty$. The
discounted bond prices $\hat{P}(t,T)$ are defined by
\begin{gather*}
\hat{P}(t,T):=e^{-\int_{0}^{t}v(s)ds}\cdot P(t,T),\qquad 0\leq t\leq
T<+\infty,
\end{gather*}
where $v(t):=f(t,t), t\geq 0$ is the short rate. If we extend the
domain of $f$ by putting $f(t,T)=f(T,T)$ for $t\geq T$ we obtain the
formula
\begin{gather*}
\hat{P}(t,T)=e^{-\int_{0}^{T}f(t,u)du}, \qquad 0\leq t\leq T< +\infty.
\end{gather*}
The market is supposed to be arbitrage free, i.e. we assume that the processes $\hat{P}(\cdot,T)$ are local martingales. This implies that the
coefficients $\alpha,\sigma$ in \eqref{rownanie na f} satisfy the Heath-Jarrow-Morton condition, i.e. for each $T\geq0$
\begin{gather}\label{warunek HJM}
\int_{t}^{T}\alpha(t,u)du=J\left(\int_{t}^{T}\sigma(t,u)du\right),
\end{gather}
for almost all $t\geq0$, see \cite{bjork}, \cite{Eberlein}, \cite{JakubowskiZabczyk}. The function $J$ above is the Laplace exponent of $L$ defined by
\eqref{J}. As $J$ is differentiable, \eqref{warunek HJM} can be written as
\begin{gather*}
\alpha(t,T)=J^\prime\left(\int_{t}^{T}\sigma(t,u)du\right)\sigma(t,T), \quad 0\leq t\leq T<+\infty,
\end{gather*}
which means that the drift is fully determined by the volatility process. As a consequence  \eqref{rownanie na f} reads as
\begin{gather}
f(t,T)=f(0,T)+\int_{0}^{t}J^{\prime}\left(\int_{s}^{T}\sigma(s,u)du\right)\sigma(s,T)ds+\int_{0}^{t}\sigma(s,T)dL(s), \quad 0\leq t\leq T<+\infty.
\end{gather}
If we put $x=T-t$ then from the above we obtain \eqref{semigroup equation} for the dynamics of $r(t,x)$, which is a weak form of \eqref{general
equation}.\vskip2mm

\noindent{The assumptions that the process $r(t,\cdot), t\geq 0$, takes values in $L^{2,\gamma}$ or in $H^{1,\gamma}$ have  financial
interpretations. For instance if  $r\in L^{2,\gamma}$ then
\begin{align}\label{calkowalnosc rozwiazania}\nonumber
\int_{0}^{+\infty}\mid r(x)\mid dx&=\int_{0}^{+\infty}\mid r(x)\mid e^{\frac{\gamma}{2}x}\cdot e^{-\frac{\gamma}{2}x}dx\leq \left(\int_{0}^{+\infty}\mid
r(x)\mid^{2}e^{\gamma x}dx\right)^{\frac{1}{2}}\left(\int_{0}^{+\infty}e^{-\gamma
x}dx\right)^\frac{1}{2}\\[2ex]
&\leq \frac{1}{\sqrt{\gamma}} \ \|r\|_{L^{2,\gamma}} <+\infty.
\end{align}
Consequently, for fixed $t$ and all \,\,$T\geq t$,
\begin{gather*}
P(t,T)\geq e^{-\,\,{\frac{1}{\sqrt{\gamma}}   \|r(t)\|_{L^{2,\gamma}}}},
\end{gather*}
and therefore the bond prices, as functions of the maturity $T$, are bounded from below by a positive number. The requirement
\begin{gather*}
\int_{0}^{+\infty}\mid r^{\prime}_{x}(t,x)\mid^2e^{\gamma x}dx<+\infty
\end{gather*}
corresponds to the observation that the forward rates are getting flat for large maturities $T$.}
\subsection{Laplace exponent}\label{Laplace exponent}

To examine properties of the Laplace exponent
\begin{gather*}
 J(z)=-az+\frac{1}{2}qz^2+\int_{-\infty}^{+\infty}(e^{-zy}-1+zy\mathbf{1}_{(-1, 1)}(y)) \ \nu(dy),\qquad z\in\mathbb{R},
\end{gather*}
let us represent it in the form
\begin{gather*}
 J(z)=-az+\frac{1}{2}qz^2+J_1(z)+J_2(z)+J_3(z)+J_4(z),
\end{gather*}
where
\begin{align*}
J_1(z)&:=\int_{-\infty}^{-1}(e^{-zy}-1)\nu(dy),&\qquad  J_2(z)&:=\int_{-1}^{0}(e^{-zy}-1+zy)\nu(dy),\\[2ex]
J_3(z)&:=\int_{0}^{1}(e^{-zy}-1+zy)\nu(dy),&\qquad J_4(z)&:=\int_{1}^{+\infty}(e^{-zy}-1)\nu(dy).
\end{align*}
If the integrals below exist then we have the following formulas for the derivatives, see for instance Lemma 8.1 and 8.2 in \cite{Rusinek}
\begin{align*}
J^{\prime}_1(z)&:=-\int_{-\infty}^{-1}ye^{-zy}\nu(dy),&\qquad  J^{\prime}_2(z)&:=\int_{-1}^{0}y(1-e^{-zy})\nu(dy),\\[2ex]
J^{\prime}_3(z)&:=\int_{0}^{1}y(1-e^{-zy})\nu(dy),&\qquad J^{\prime}_4(z)&:=-\int_{1}^{+\infty}ye^{-zy}\nu(dy);
\end{align*}
\begin{align*}
J^{\prime\prime}_1(z)&:=\int_{-\infty}^{-1}y^2e^{-zy}\nu(dy),&\qquad  J^{\prime\prime}_2(z)&:=\int_{-1}^{0}y^2e^{-zy}\nu(dy),\\[2ex]
J^{\prime\prime}_3(z)&:=\int_{0}^{1}y^2e^{-zy}\nu(dy),&\qquad J^{\prime\prime}_4(z)&:=\int_{1}^{+\infty}y^2e^{-zy}\nu(dy);
\end{align*}
\begin{align*}
J^{\prime\prime\prime}_1(z)&:=-\int_{-\infty}^{-1}y^3e^{-zy}\nu(dy),&\qquad  J^{\prime\prime\prime}_2(z)&:=-\int_{-1}^{0}y^3e^{-zy}\nu(dy),\\[2ex]
J^{\prime\prime\prime}_3(z)&:=-\int_{0}^{1}y^3e^{-zy}\nu(dy),&\qquad J^{\prime\prime\prime}_4(z)&:=-\int_{1}^{+\infty}y^3e^{-zy}\nu(dy).
\end{align*}
Below we gather properties of $J$ needed in the paper. The domain of
$J$ is restricted to the half-line $[0,+\infty)$ due to the fact
that we are interested in positive solutions of \eqref{general
equation} only. For $z>0, \mid J^{\prime}(z)\mid<+\infty$ if
$J_1^{\prime}(z)$ is well defined, that is if
\begin{gather*}
\int_{-\infty}^{-1}\mid y\mid e^{z\mid y\mid}\nu(dy)<+\infty.
\end{gather*}
Moreover,
\begin{itemize}
\item $\mid J^{\prime}(0)\mid<+\infty$  iff
\begin{gather} \label{J^prim w zerze skonczone}
\text{(B0)}\qquad\qquad \int_{\mid y\mid>1}\mid
y\mid\nu(dy)<+\infty,
\end{gather}
and
\item $J^{\prime}$ is increasing.
\end{itemize}
Moreover, it follows from the below formulas
\begin{gather*}
\lim_{z\rightarrow+\infty}\mid J^{\prime}_1(z)\mid=+\infty, \qquad \lim_{z\rightarrow+\infty}\mid J^{\prime}_2(z)\mid=+\infty, \\[2ex]
\mid J^{\prime}_3\mid \text{is bounded} \Longleftrightarrow \int_{0}^{1}y\nu(dy)<+\infty, \quad
 \mid J^{\prime}_4(z)\mid \text{is bounded} \Longleftrightarrow \int_{1}^{+\infty}y\nu(dy)<+\infty,
\end{gather*}
that under \eqref{J^prim w zerze skonczone}
\begin{itemize}
\item $J^{\prime}$ is bounded on $[0,+\infty)$ iff

\[ (B1) \quad
\begin{cases}
  \quad \bullet \quad L  \ \text{does not have the Wiener part, i.e.} \ q=0,\\[2ex]
   \quad \bullet \quad supp\{\nu\}\subseteq [0,+\infty),\\[2ex]
 \quad \bullet \quad \int_{0}^{+\infty}y\nu(dy)<+\infty.
 \end{cases}
\]
\end{itemize}
By similar analysis
\begin{gather*}
\lim_{z\rightarrow+\infty}\mid J^{\prime\prime}_1(z)\mid=+\infty, \qquad \lim_{z\rightarrow+\infty}\mid J^{\prime\prime}_2(z)\mid=+\infty, \\[2ex]
\mid J^{\prime\prime}_3\mid \text{is bounded}, \quad
 \mid J^{\prime\prime}_4(z)\mid \text{is bounded}\Longleftrightarrow \int_{1}^{+\infty}y^2\nu(dy)<+\infty,
\end{gather*}
we conclude that
\begin{itemize}
\item $J^{\prime\prime}$ is bounded on $[0,+\infty)$ iff
\[ (B2) \quad
\begin{cases}
  \quad \bullet \quad supp\{\nu\}\subseteq [0,+\infty),\\[2ex]
   \quad \bullet \quad \int_{1}^{+\infty}y^2\nu(dy)<+\infty.\\[2ex]
  \end{cases}
\]
\end{itemize}
\noindent
$J^{\prime}$ is bounded on $[0,z_0], z_0>0$ iff $
\Big\vert\int_{\mid y\mid>1}ye^{-z_0y}\nu(dy)\Big\vert<+\infty
$
and is finite at $0$. Thus
\begin{itemize}
\item $J^{\prime}$ is locally bounded iff
 \eqref{J^prim w zerze skonczone} holds and $\int_{y<-1}\mid y\mid e^{z_0\mid y\mid}\nu(dy)<+\infty$.
\end{itemize}
Similarly,
\begin{itemize}
 \item $J^{\prime}$ is locally Lipschitz iff
\[ (L1) \quad
\begin{cases}
  \quad \bullet \quad \int_{-\infty}^{-1} \mid y\mid^2 e^{z_0\mid y\mid}\nu(dy)<+\infty,\\[2ex]
   \quad \bullet \quad \int_{\mid y\mid>1}\mid y\mid^2\nu(dy)<+\infty,
 \end{cases}
\]
\end{itemize}
and
\begin{itemize}
 \item $J^{\prime\prime}$ is locally Lipschitz iff
 \[ (L2) \quad
\begin{cases}
  \quad \bullet \quad \int_{-\infty}^{-1} \mid y\mid^3 e^{z_0\mid y\mid}\nu(dy)<+\infty,\\[2ex]
   \quad \bullet \quad \int_{\mid y\mid>1}\mid y\mid^3\nu(dy)<+\infty.
 \end{cases}
\]
\end{itemize}

For the linear case we assume that the support of the L\'evy measure is contained in $[-\frac{1}{\bar{\lambda}}, +\infty)$, where $-\infty<\bar{\lambda}<+\infty$.  Thus the above results can be written in the simpler form.
The assumption \eqref{J^prim w zerze skonczone} reduces to the form
\begin{itemize}
\item
\begin{gather}\label{J^prim w zerze skonczone liniowy przypadek}
\mid J^{\prime}(0)\mid<+\infty \quad \Longleftrightarrow \quad \int_{y >1} y \nu(dy)<+\infty,
\end{gather}
\end{itemize}
and
 \begin{itemize}
 \item $J^{\prime}$ is locally bounded $\Longleftrightarrow$ \eqref{J^prim w zerze skonczone liniowy przypadek} holds,
 \item $J^{\prime}$ is locally Lipschitz $\Longleftrightarrow$ $\int_{1}^{+\infty}y^2\nu(dy)<+\infty$,
 \item $J^{\prime\prime}$ is locally Lipschitz $\Longleftrightarrow$ $\int_{1}^{+\infty}y^3\nu(dy)<+\infty$.
 \end{itemize}

\end{document}